\documentclass[a4paper,prr,twocolumn]{revtex4-2}

\usepackage{amsfonts}
\usepackage{amsmath,amssymb}
\usepackage{amsthm}
\usepackage{graphicx}
\usepackage{physics}
\usepackage[colorlinks,citecolor=DarkGreen,linkcolor=FireBrick]{hyperref}

\usepackage{multirow}

\usepackage[svgnames,psnames]{xcolor}

\newcommand{\Kcpl}{{K_\mathrm{cpl}}}
\newcommand{\ND}{{N_\mathrm{q}}}

\usepackage{orcidlink}

\usepackage{comment}

\newcommand{\cD}{{\mathcal{D}}}

\newcommand{\cH}{{\mathcal{H}}}

\newcommand{\cO}{{\mathcal{O}}}

\begin{document}

\begin{titlepage}
\thispagestyle{empty}

\begin{flushright}
YITP-23-31 
\end{flushright}

\title{Hayden-Preskill Recovery in Hamiltonian Systems}
\author{Yoshifumi Nakata$^{1}$ \orcidlink{0000-0003-1285-6968}}
\email{yoshifumi.nakata@yukawa.kyoto-u.ac.jp}
\author{Masaki Tezuka$^{2}$ \orcidlink{0000-0001-7877-0839}}
\email{tezuka@scphys.kyoto-u.ac.jp}
\affiliation{$^1$Yukawa Institute for Theoretical Physics, Kyoto University, Kitashirakawa, Sakyo-ku, Kyoto 606-8502, Japan,}
\affiliation{$^2$Department of Physics, Kyoto University, Kitashirakawa, Sakyo-ku, Kyoto 606-8502, Japan.}
\date{\today}

\begin{abstract}
Information scrambling refers to the unitary dynamics that quickly spreads and encodes localized quantum information over an entire many-body system and makes the information accessible from any small subsystem. While information scrambling is the key to understanding complex quantum many-body dynamics and is well-understood in random unitary models, it has been hardly explored in Hamiltonian systems. In this Letter, we investigate the information recovery in various time-independent Hamiltonian systems, including chaotic spin chains and Sachdev-Ye-Kitaev (SYK) models. We show that information recovery is possible in certain, but not all, chaotic models, which highlights the difference between information recovery and quantum chaos based on the energy spectrum or the out-of-time-ordered correlators. We also show that information recovery probes transitions caused by the change of information-theoretic features of the dynamics.
\end{abstract}
\maketitle
\end{titlepage}

\emph{Introduction.}
A central challenge in modern physics is to characterize the dynamics in far-from-equilibrium quantum systems. The Hayden-Preskill protocol~\cite{HP2007} offers an operational approach toward this goal and has been attracting much attention~\cite{SS2008,S2011,LSHOH2013,SS2014,SS2015,RS2015,RY2017,Y2019,L2020,Kitaev1,Kitaev2,Jensen2016PhysRevLett.117.111601,Maldacena_Stanford2016PhysRevD.94.106002,Sachdev2015,B2016,vKRPS2018,KVH2018,Hosur_2016,PYHP2015,PEW2017,KC2019,HP2019,NWK2022,L2020,NMK2022,CLGCZZ2020,LFSLYYM2019,BGLLNSSSW2021,NLBGLSSSW2021}. The protocol addresses if the information initially localized in a small subsystem can be recovered from other subsystems after unitary time evolution. 
If the unitary dynamics is sufficiently random, the information is rapidly encoded into the whole system, and information can be recovered from any small subsystem~\cite{HP2007}. This phenomenon is can be thought of an information-theoretic manifestation of complex quantum dynamics and is called the \emph{Hayden-Preskill recovery}. 
The unitary dynamics that leads to the Hayden-Preskill recovery is referred to as \emph{information scrambling}~\cite{HP2007}.

The Hayden-Preskill recovery is of interdisciplinary interest: it is inspired by the information paradox of black holes~\cite{hawking_black_1974, hawking_particle_1975, Hawking1976PhysRevD.14.2460}, is formulated in the quantum information language, and is investigated by the technique of random matrix theory (RMT)~\cite{Haake-2001,mehta2004random}.
Many related properties, such as entanglement generation~\cite{SS2008,LSHOH2013}, operator mutual information (OMI)~\cite{Hosur_2016,caputa_scrambling_2015,Nie_2019,https://doi.org/10.48550/arxiv.2302.08009}, and out-of-time-ordered correlators (OTOCs)~\cite{Larkin-Ovchinnikov1969,Shen-Zhang-Fan-Zhai2017,CLGCZZ2020}, have been intensely studied.

Despite these progresses, information recovery in Hamiltonian systems has been rarely explored~\cite{https://doi.org/10.48550/arxiv.2301.07108}.
It is widely believed that the Hayden-Preskill recovery is possible in quantum chaotic systems, but the original analysis strongly relies on the random unitary assumption, which is unlikely to be satisfied even approximately by time-independent Hamiltonian dynamics~\cite{RY2017, cotler_chaos_2017}. 
Furthermore, quantum chaos is commonly characterized by eigenenergy statistics, which is a \emph{static} property, but information recovery is about the \emph{dynamical} properties. Thus, the relation between quantum chaos and information recovery is not a priori trivial.

In this Letter, we investigate in detail the information recovery in various Hamiltonian systems. We first provide a class of Hamiltonians that do not lead to information scrambling. This includes chaotic spin-$1/2$ chains, such as the Heisenberg model with random magnetic field and the mixed field Ising model. Notably, they saturate OTOCs for local observables but do not achieve the Hayden-Preskill recovery, demonstrating the difference between the saturation of OTOCs for local observables and information scrambling.

We then confirm information scrambling in the Sachdev-Ye-Kitaev (SYK) Hamiltonian~\cite{Sachdev1993PhysRevLett.70.3339,Sachdev2010PhysRevLett.105.151602,Kitaev1,Kitaev2}, which is a canonical holographic dual to quantum gravity~\cite{Maldacena_Stanford2016PhysRevD.94.106002,Jensen2016PhysRevLett.117.111601,Cotler_2017,Sarosi_2018,Trunin_2021}. The SYK model is known to have `scrambling' features in many senses, such as saturation of OTOCs~\cite{Kitaev1,Kitaev2} for local observables, the maximum quantum Lyapunov exponent~\cite{MSSbound}, and an RMT-like energy statistics~\cite{You_2017,Garcia_Garcia_2016,Cotler_2017}. 
Our result adds another scrambling feature to the model, that is, it achieves the Hayden-Preskill recovery.
We also show that sparse variants~\cite{SachdevReview2022} achieve the information recovery as well, possibly helping experimental realizations of the protocol.

We finally address the question whether the information recovery can reveal novel quantum many-body phenomena. Using a variant of SYK models, we affirmatively answer to this question: the information recovery can capture a transition that was previously unknown. The transition is caused by a drastic change of information-theoretic structures of the Hamiltonian dynamics. This is of interest as it characterizes complex quantum many-body dynamics from the quantum information viewpoint.

\emph{The Hayden-Preskill protocol.}
Given quantum many-body system $S$ of $N$ qubits, we encode quantum information into a local subsystem $A \subseteq S$ of $k$ qubits ($k \ll N$). We then let the system $S$ undergo the Hamiltonian time-evolution $U_{\hat{H}}(t) := e^{-i \hat{H} t}$ for some time $t$, where $\hat{H}$ is the Hamiltonian in $S$. After the time-evolution, the information is tried to be recovered from an $\ell$-qubit subsystem $C \subseteq S$. Throughout our analysis, we assume $C \subseteq B := S \setminus A$ as far as $\ell \leq N-k$. The question is how large $\ell$ should be for a successful recovery. 

\begin{figure}
    \centering
    \includegraphics[width=0.5\linewidth]{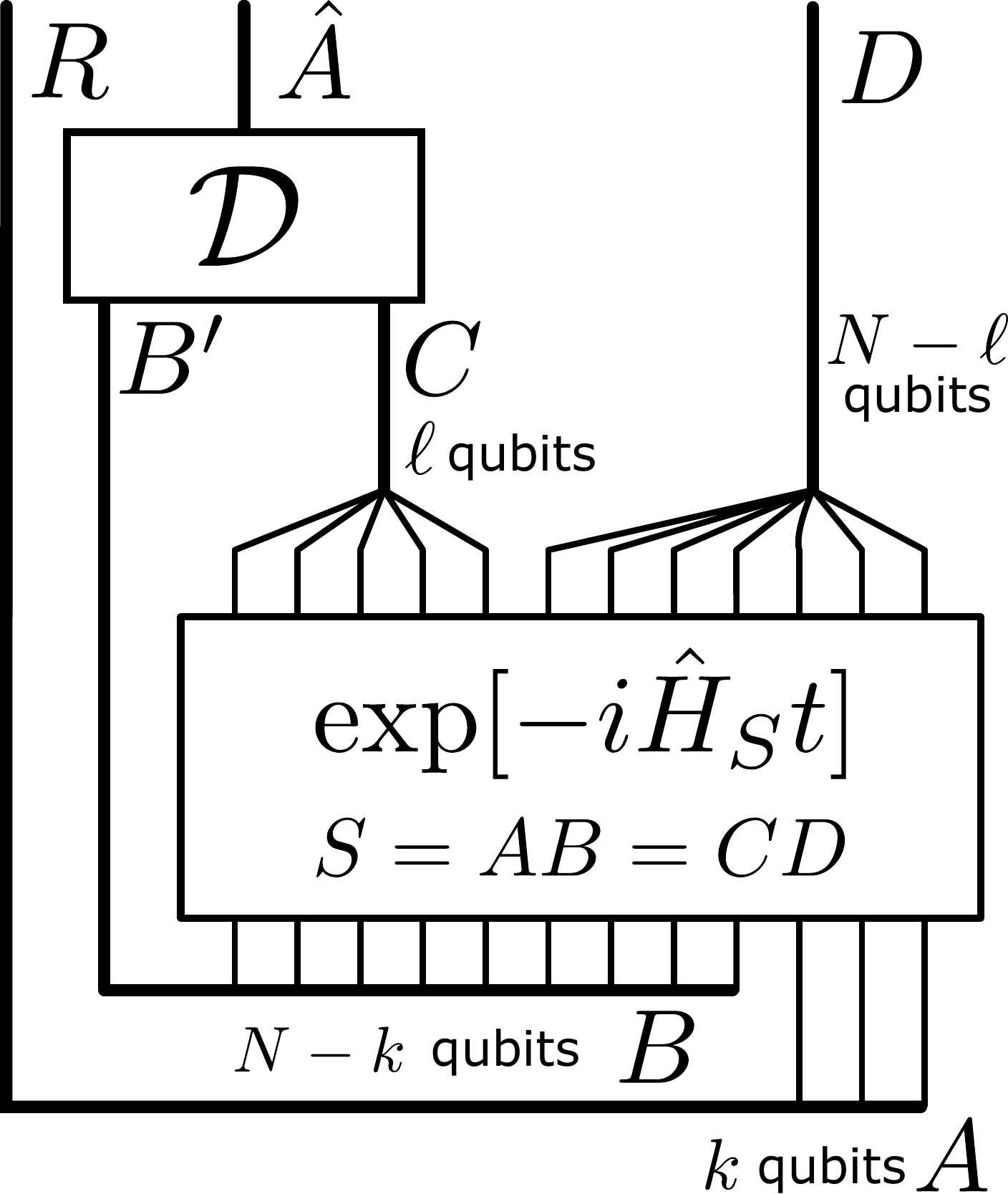}
    \caption{A diagram of the Hayden-Preskill protocol. Time flows from bottom to top. Horizontal lines imply that the qubits connected by the line may be entangled. The initial states on $AR$ and $BB'$ are given by a maximally entangled state of $k$ ebits, i.e., $k$ EPR pairs, that keeps track of quantum information in $A$, and a purified state $\ket{\xi(\beta)}$ of a thermal state on $B$ at the inverse temperature $\beta$, respectively. The system $S:=AB$ undergoes Hamiltonian dynamics by $\hat{H}_S$, and then, is split into two subsystems, $C$ of $\ell$ qubits and $D$ of $N-\ell$ qubits. By applying a quantum channel $\cD$ onto $B'C$, one aims to decode the quantum information in $A$, that is, to recover the $k$ EPR pairs between $\hat{A}$ and $R$. This protocol has a natural interpretation in the context of information paradox~\cite{HP2007}. See also a tutorial~\cite{Xu-Swingle-tutorial2022}.}
    \label{fig:Hayden-Preskillsetting}
\end{figure}

The answer depends on the initial state in $B$ as well as available resources in the recovery process. Here, we assume that the initial state in $B$ is a thermal state $\xi^B(\beta)$ at inverse temperature $\beta$. Based on its eigendecomposition $\xi^B(\beta) = \sum_j p_j(\beta) \ketbra{\psi_j}{\psi_j}^B$, we introduce a purified state $\ket{\xi(\beta)}^{B B'} := \sum_j \sqrt{p_j(\beta)} \ket{\psi_j}^B \otimes \ket{\psi_j}^{B'}$ on the system $BB'$. When the subsystem $B'$ is traced out, the marginal state on $B$ is the original thermal state $\xi^B(\beta)$.
We consider the scenario in which the subsystem $B'$ can be used in the recovery process. This has a natural interpretation in the black hole information paradox~\cite{HP2007} and is of considerable interest. See Fig.~\ref{fig:Hayden-Preskillsetting}.

The recovery of quantum information is formally defined by introducing a virtual reference system $R$ that keeps track of the quantum information. Denoting $k$ Einstein--Podolsky--Rosen (EPR) pairs bertween $A$ and $R$ by $\ket{\Phi}^{AR}$, we set the initial state on the system $SB'R = ABB'R$ to $\ket{\Psi(t=0,\beta)}^{S B'R} = \ket{\Phi}^{AR} \otimes \ket{\xi(\beta)}^{B B'}$. The subsystem $S$ undergoes the Hamiltonian time-evolution by $\hat{H}_S$, resulting in the state $\ket{\Psi(t, \beta)}^{S B' R}=(I^{B'R} \otimes e^{-i \hat{H}_S t}) \ket{\Psi(t, \beta)}^{S B'R}$ at time $t$. 
Let $\Psi^{C B' R}(t, \beta)$ be the marginal state on $CB'R$, which is given by taking the partial trace over $D$ of $\ket{\Psi(t, \beta)}^{S B' R}$. In the following, we indicate the subsystem over which the partial trace is taken by omitting the subsystem from the superscript. 

Following the standard convention~\cite{HP2007}, the recovery error is defined by
\begin{equation}
    \Delta_{\hat{H}}(t, \beta) := \frac{1}{2} \min_{\cD} 
    \norm{
    \ketbra{\Phi}{\Phi}^{AR} - \cD^{C B' \rightarrow A} \bigl( \Psi^{C B' R}(t, \beta) \bigr) 
    }_1.
\end{equation}
Here, the minimum is taken over all possible quantum operations $\cD$, namely, all completely-positive and trace-preserving (CPTP) maps, from $C B'$ to $A$, and $\|\rho\|_1 = \mathrm{Tr}\sqrt{\rho^{\dagger}\rho}$ is the trace norm. Due to the Holevo-Helstrom theorem~\cite{NC2000}, the trace norm between two quantum states characterizes how well they can be distinguished and is suitable to quantify the recovery error.
We normalize $\Delta_{\hat{H}}(t, \beta)$ so that $0 \leq \Delta_{\hat{H}}(t, \beta) \leq 1$. See \ref{S:bounds} \cite{Supplemental}.

Computing $\Delta_{\hat{H}}(t, \beta)$ is in general intractable due to the minimization over the CPTP maps. The \emph{decoupling} condition provides a necessary and sufficient condition for the recovery in terms of the state $\Psi^{D R}(t, \beta)$~\cite{doi:10.1142/S1230161208000043,dupuis_one-shot_2014,Dupuis1004.1641}. Using the condition, a calculable, and typically good, upper bound on $\Delta_{\hat{H}}(t, \beta)$ is obtained:
\begin{equation}
    \Delta_{\hat{H}}(t, \beta) 
    \leq 
    \bar{\Delta}_{\hat{H}}(t, \beta)
    := 
    \min \bigl\{1, \sqrt{2\Sigma_{\hat{H}}(t, \beta)} \bigr\}, \label{Eq:UpperBound}
\end{equation}
where $\Sigma_{\hat{H}}(t, \beta) := \|\Psi^{D R}(t, \beta)-\Psi^{D}(t, \beta) \otimes \pi^R \|_1/2$, and $\pi^R = I^R/2^k$ is the completely mixed state on $R$. See also \ref{S:bounds} \cite{Supplemental}.
Note that the quantity $\Sigma_{\hat{H}}(t, \beta)$ is closely related to the mutual information between $R$ and $D$, which is equivalent to the OMI~\cite{Hosur_2016}, but the OMI leads to a worse bound than Eq.~\eqref{Eq:UpperBound}.

The recovery error $\Delta_{\hat{H}}(t, \beta)$ is also related to OTOCs. If one can compute OTOCs for all observables on the $k$-qubit subsystem $A$ and the $\ell$-qubit subsystem $C$, or all the $4^{k + \ell}$ operators that form an operator basis on $AC$, the recovery error could be evaluated~\cite{yoshida2017efficient,PhysRevX.9.011006}. 
However, this is computationally intractable as OTOCs for at least $4^{k + \ell}$ operators are needed.
Note that the existing studies of OTOCs in Hamiltonian systems are mostly about the cases with $k=\ell=1$, and hence, do not provide much insight into the recovery error.

\emph{Random unitary model and information scrambling.}
The Hayden-Preskill protocol was understood well in a random unitary model, where the time evolution $e^{-i \hat{H_S}t}$ is replaced with a Haar random unitary.
The model does not have a parameter corresponding to time $t$, and its recovery error $\Delta_{\rm Haar}(\beta)$ satisfies, with high probability,
\begin{equation}
    \Delta_{\rm Haar}(\beta) 
    \leq 
    \bar{\Delta}_{\rm Haar}(\beta)
    :=
    \min \{1, 2^{\frac{1}{2} ( \ell_{\rm Haar, th}(\beta) - \ell)} \},\label{Eq:Haar}
\end{equation}
where $\ell_{\rm Haar, th}(\beta)  := \frac{1}{2} \bigl( N+k - H(\beta) \bigr)$, and $H(\beta) = -\log[\Tr\bigl[ \bigl(\xi^B(\beta) \bigl)^2 \bigr]]$ is the Renyi-$2$ entropy~\cite{HP2007,Dupuis1004.1641,NWK2022}.

From Eq.~\eqref{Eq:Haar}, $\Delta_{\rm Haar}(\beta) \ll 1$ if $\ell \gg \ell_{\rm Haar, th}(\beta)$. In particular, $\ell_{\rm Haar, th}(0) = k$. Hence, if the system $B$ is initially at infinite temperature, the $k$-qubit quantum information in $A$ is recoverable with exponential precision from any subsystem of size larger than $k$, which is independent of $N$. This phenomena is referred to as the Hayden-Preskill recovery.
Following the original proposal~\cite{HP2007}, we refer to the dynamics achieving the Hayden-Preskill recovery as information scrambling.

\emph{Hamiltonians without information scrambling.}
Due to the facts that the information scrambling occurs in the random unitary model and that quantum chaos can be characterized by RMT, information scrambling has been commonly studied in relation with quantum chaos. However, information scrambling is not necessarily related to the quantum chaos in terms of the RMT-like energy statistics.
For instance, we can analytically show that the dynamics of any commuting Hamiltonians is not information scrambling (see \ref{S:CH} \cite{Supplemental}), while they can have RMT-like features~\cite{PhysRevLett.98.017201, gur-ari_does_2018} in the energy spectrum. 

\begin{figure}
    \includegraphics[width=\linewidth]{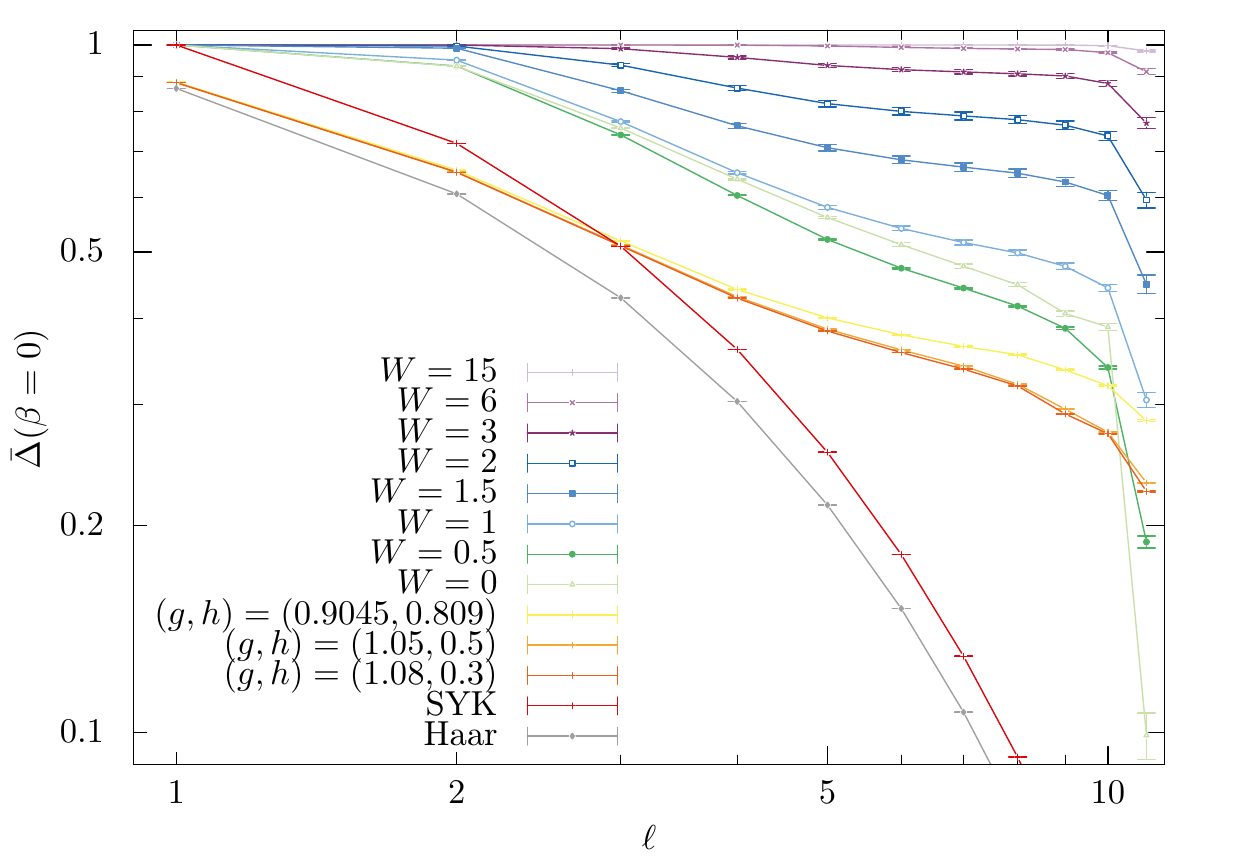}
    \caption{
    Semilogarithmic plot of the late-time values of $\bar{\Delta}$ for $\hat{H}_\mathrm{XXZ}$, $\hat{H}_{\rm Ising}$, and $\hat{H}_{\mathrm{SYK}_4}$.
    To compare, the values for the Haar random unitary are also plotted.
    $k=1$ for all models, and $N=12$ ($\ND=12$) is chosen for the XXZ and Ising (SYK$_4$) models. The dimension of the Haar random unitary is $2^{12}$.
    Note that the conservation of the $z$ component of spins (parity) is considered for the Heisenberg spin chain (SYK model).
    The values of $(g,h)$ are those discussed in \cite{rodrigueznieva2023quantifying}.
    The averages for $t=(1,2,.\ldots,10)\times10^6$ are plotted as the late-time value. For random-field average, 16 samples are taken.}
    \label{fig:XXZ-Ising-L12}
\end{figure}

More illuminating instances are the spin-$1/2$ chains such as the Heisenberg with random magnetic field, $\hat{H}_\mathrm{XXZ}= \frac{1}{4}\sum_{j=1}^{N-1}\left(X_j X_{j+1} + Y_j Y_{j+1} + J_z Z_{j} Z_{j+1}\right)+ \frac{1}{2} \sum_{j=1}^N h_j Z_j$, where $h_j$ are independently sampled from a uniform distribution in $[-W,W]$, and the mixed-field Ising with constant magnetic field, $\hat{H}_\mathrm{Ising}
= -\sum_{j=1}^{N-1} \left(Z_{j} Z_{j+1}\right) -g\sum_{j=1}^N X_j-h\sum_{j=1}^N Z_j$. Here, $X_j, Y_j$, and $Z_j$ denote the Pauli matrices on site $j$. 
In contrast to the fact that both have integrable--chaotic transitions by varying parameters~\cite{LFSantos_2004, Kudo_Deguchi_PhysRevB.69.132404,Viola_Brown_JPhysA_2007,Znidaric_Prosen_Prelovsek_PhysRevB.77.064426,Pal_Huse_PhysRevB.82.174411.2010,De_Luca_Scardicchio_EPL_2013,Alet_Laflorencie_2018,RevModPhys.91.021001,Luitz-Laflorencie-Alet-PhysRevB.91.081103,Morningstar_Huse_PRB2019,Suntajs_Bonca_Prosen_Vidmar_PhysRevE.102.062144,Sierant_Delande_Zakrzewski_2020,Kiefer-Emmanouilidis-PhysRevLett.124.243601,Chanda_Sierant_Zakrzewski_PhysRevResearch.2.032045,Sels-Polkovnikov-PhysRevE.104.054105,Kiefer-Emmanouilidis-PhysRevB.103.024203,Morningstar__PhysRevB.105.174205,Sels_PhysRevB.106.L020202,Sierant_Zakrzewski_PhysRevB.105.224203,Ghosh_Znidaric_PRB2022,Banuls2011,rodrigueznieva2023quantifying}, our numerical analysis reveals that the recovery errors are not as small as the random unitary model for any values of parameters at any time $t$. 

This is demonstrated in Fig.~\ref{fig:XXZ-Ising-L12}, where the late-time values of $\bar{\Delta}_{\hat{H}}(t, \beta=0)$ are plotted for these Hamiltonians. 
Hereafter, we set $k$ to $1$ in all numerics throughout the paper for the sake of computational tractability.
It is observed that, by increasing $\ell$, $\bar{\Delta}_{\hat{H}}$ decays inverse-polynomially or more slowly. We also provide in \ref{S:Heisenberg} and \ref{S:Ising} \cite{Supplemental} evidences that these values do not depend on the system size $N$. This implies that, although Fig.~\ref{fig:XXZ-Ising-L12} is for $N = 12$ and $\ell \leq N-1 = 11$, we can infer $\bar{\Delta}_{\hat{H}}$ for larger $N$ and $\ell$ by extrapolation. By doing so, we may observe that $\bar{\Delta}_{\hat{H}}$ for $W \gtrsim 1$ may possibly stay nearly constant in the large-$N$ limit unless $\ell \approx N$.

\begin{figure}
    \includegraphics[width=\linewidth]{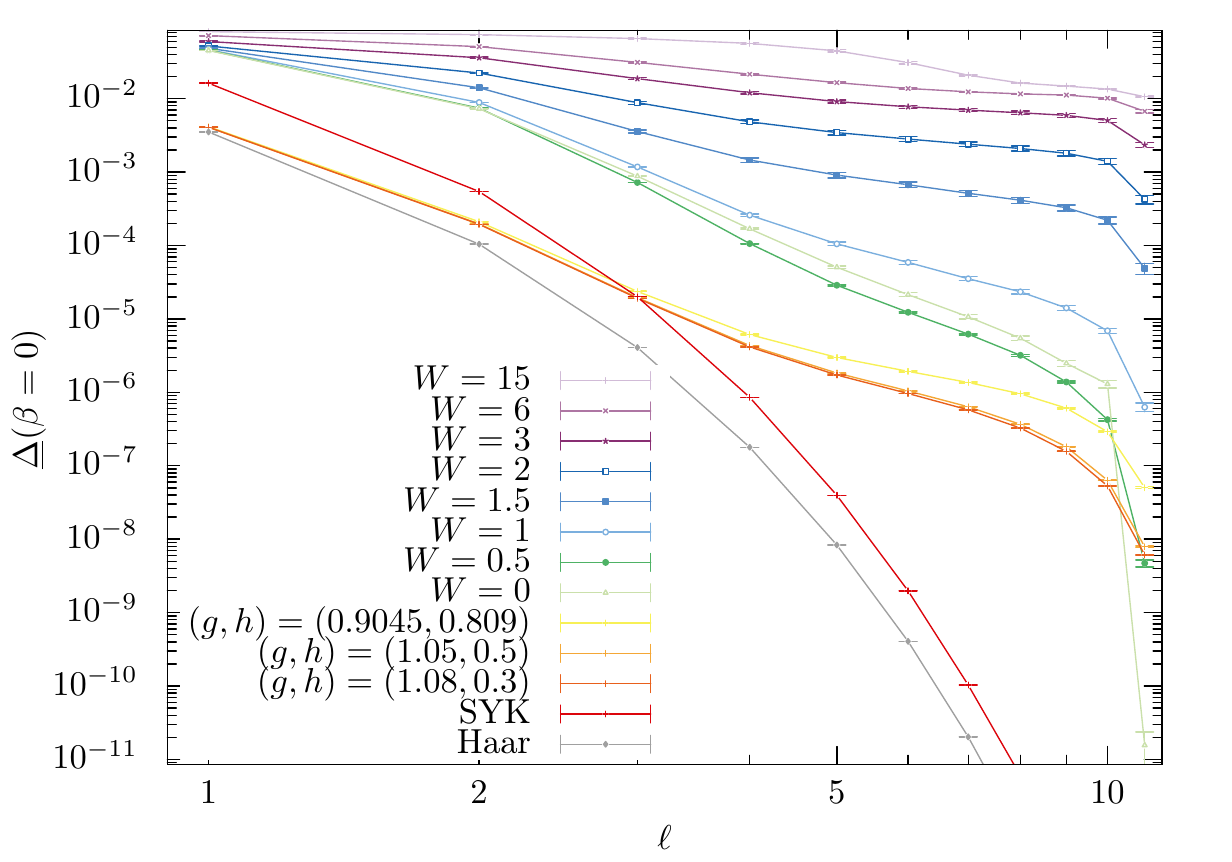}
    \caption{
    Semilogarithmic plot of the late-time values of $\underline{\Delta}$ for $\hat{H}_\mathrm{XXZ}$, $\hat{H}_{\rm Ising}$ ($N=12$), and $\hat{H}_{\mathrm{SYK}_4}$ ($\ND=12$). For the Hamiltonians with random field, averages over $16$ samples are taken.}
    \label{fig:IRBC}
\end{figure}

We can also investigate lower bounds on the recovery errors based on the mutual information, which we denote by $\underline{\Delta}_{\hat{H}}$ (see~\ref{SS:LBbyMI} \cite{Supplemental} for the derivation).
The bound is not tight, but we show in Fig.~\ref{fig:IRBC} that the lower bounds for $\hat{H}_\mathrm{XXZ}$ and $\hat{H}_\mathrm{Ising}$ scale similarly to those of the upper bounds. That is, they decay inverse-polynomially or more slowly as $\ell$ increases and possibly stay almost constant if $W \gtrsim 1$ unless $\ell \approx N$. 

As both upper and lower bounds scale similarly, we reasonably conclude $\Delta_{\hat{H}} = \Omega(1/{\rm poly}(\ell))$ in the large-$N$ limit. This is in sharp contrast to the exponential decay of the recovery error in the random unitary model and implies that the dynamics of these Hamiltonians is not information scrambling in any parameter region. 

More closely looking at Figs.~\ref{fig:XXZ-Ising-L12} and~\ref{fig:IRBC}, $\Delta_{\hat{H}}$ is likely to be dependent on the parameters of the Hamiltonians. It is known that $\hat{H}_\mathrm{XXZ}$ shows integrable-chaotic-MBL transitions as $W$ increases and that the system is chaotic for $W \approx 0.5$~\cite{Pal_Huse_PhysRevB.82.174411.2010,Luitz-Laflorencie-Alet-PhysRevB.91.081103}. However, this chaotic transition does not seem to have strong consequence to information scrambling as both upper and lower bounds on $\Delta_{\hat{H}}$ for $W=0.5$ are only slightly smaller than that in the integrable case with $W=0$. 
For $\hat{H}_\mathrm{Ising}$, while the parameter $(g, h) = (1.08, 0.3)$ leads to the most chaotic feature in the entanglement structure~\cite{rodrigueznieva2023quantifying}, both upper and lower bounds on $\Delta_{\hat{H}}$ for that value can be worse than other parameters. This also indicates that information scrambling differs from quantum chaos and may not be able to be inferred from static features of Hamiltonians.

The fact that information scrambling is not observed in these systems does not contradict to the saturation of OTOCs for local, typically single-qubit, observables at late time when the parameters are appropriately set~\cite{PhysRevX.7.031011,Fan-Zhang-Shen-Zhai2017,PhysRevB.99.054205,PhysRevB.99.184202,s41567-019-0712-4,PhysRevB.105.224307}. Our numerital results rather indicate that OTOCs for multi-qubit observables are not saturated in such cases~\cite{yoshida2017efficient,PhysRevX.9.011006}, which may be of independent interest.

\emph{Original and sparse SYK Hamiltonians.}
From these results, it is likely that more drastic Hamiltonians are needed to achieve the information scrambling. We next consider the SYK model, $\mathrm{SYK}_4$, of $2\ND$ Majorana fermions:
\begin{equation}
    \hat{H}_{\mathrm{SYK}_4} = \sum_{1\leq a_1<a_2<a_3<a_4\leq2\ND} J_{a_1a_2a_3a_4} \hat{\psi}_{a_1}\hat{\psi}_{a_2}\hat{\psi}_{a_3}\hat{\psi}_{a_4},\label{eqn:SYK4}
\end{equation}
with $\hat{\psi}_j$ being Majorana fermion operators. The couplings $J_{a_1a_2a_3a_4}$ are independently chosen at random from the Gaussian with average zero and $\sigma^2= \binom{2\ND}{4}^{-1}$.
Since the parity symmetry of $\mathrm{SYK}_4$ leads to deviations in the information recovery~\cite{L2020,Y2019,NWK2022,https://doi.org/10.48550/arxiv.2103.01876}, we focus on the even-parity sector and set $N=\ND-1$. The recovery error of the corresponding random unitary model is given by 
$\bar{\Delta}'_{\rm Haar}(\beta)=\min\{1,2^{(\ell_{\rm Haar,th}(\beta)-\ell)-\frac{1}{4}}\}$. 
See~\ref{S:pureSYK} \cite{Supplemental} for details. 
We have also checked that the effect by the periodicity, characterized by $\ND \mod 4$, is negligible.

\begin{figure}
    \centering
    \includegraphics[width=\linewidth]{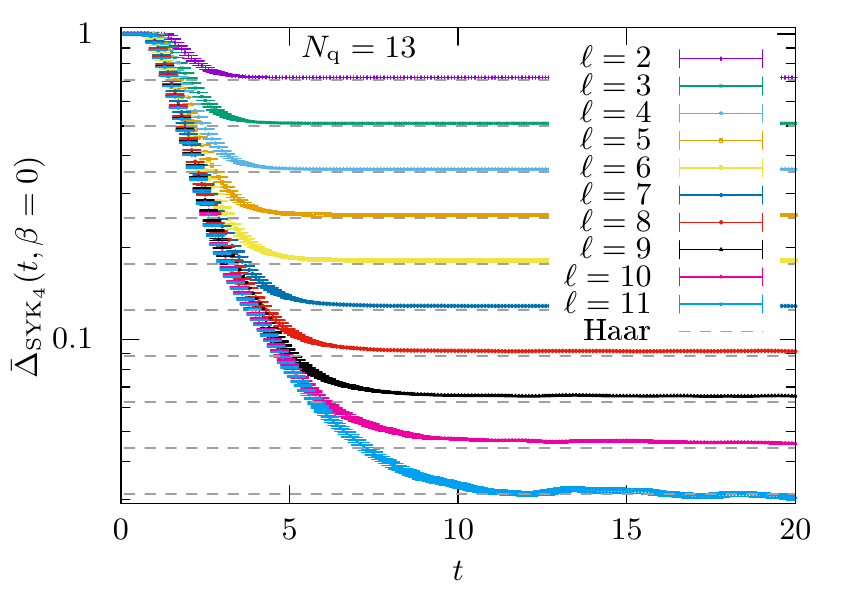}
    \caption{
    Semilogarithmic plot of the value of $\bar{\Delta}_{\hat{H}_{\mathrm{SYK}_4}}(t,\beta=0)$ against $t$ for $\ND=13$, and $2\leq \ell < \ND-k$.
    Average over 64 samples is taken.
    For $\ell=1$, $\bar{\Delta}_{\hat{H}_{\mathrm{SYK}_4}}(t,\beta=0)$ is $\sim 1$ for all $t$.
    The dashed lines represent $\bar{\Delta}'_{\rm Haar}(\beta = 0)=2^{\frac{1}{2}(1 - \ell)}$ given in Eq.~\eqref{Eq:Haar} for $\ell=2,3,\ldots,\ND-2$.
    For smaller values of $\ND$ and finite $\beta$, see \ref{S:pureSYK} \cite{Supplemental}.
    }
    \label{fig:k1N13}
\end{figure}

In Fig.~\ref{fig:k1N13}, we numerically plot the upper bound on the recovery error, $\bar{\Delta}_{\mathrm{SYK}_4}(t, \beta=0)$, against time $t$ for various $\ell$. It clearly shows that $\bar{\Delta}_{\mathrm{SYK}_4}$ quickly approaches $\bar{\Delta}'_{\rm Haar}$. This is also the case for $\beta >0$. 
We estimate that $\bar{\Delta}_{\mathrm{SYK}_4}$ converges to $\bar{\Delta}'_{\rm Haar}$ before time $t=\cO(\sqrt{\ND})$,
which qualitatively supports the fast scrambling conjecture~\cite{SS2008,S2011,LSHOH2013}.
Hence, the $\mathrm{SYK}_4$ dynamics, while differs from Haar random, has an excellent agreement with the prediction by RMT and achieves the Hayden-Preskill recovery. 
See also Figs.~\ref{fig:IRBC} and~\ref{fig:XXZ-Ising-L12}.

The situation remains the same even for a sparse simplification of $\mathrm{SYK}_4$, $\mathrm{spSYK}_4$.
In $\mathrm{spSYK}_4$, the number of non-zero random coupling constant is fixed to $\Kcpl$. It recovers $\mathrm{SYK}_4$ when $\Kcpl=\binom{2\ND}{4}$, but $\Kcpl = \cO(\ND)$ is known to suffice to have chaotic features and to reproduce holographic properties~\cite{xu2020sparse,Garc_a_Garc_a_2021}. 

\begin{figure}
    \centering
    \includegraphics[width=\linewidth]{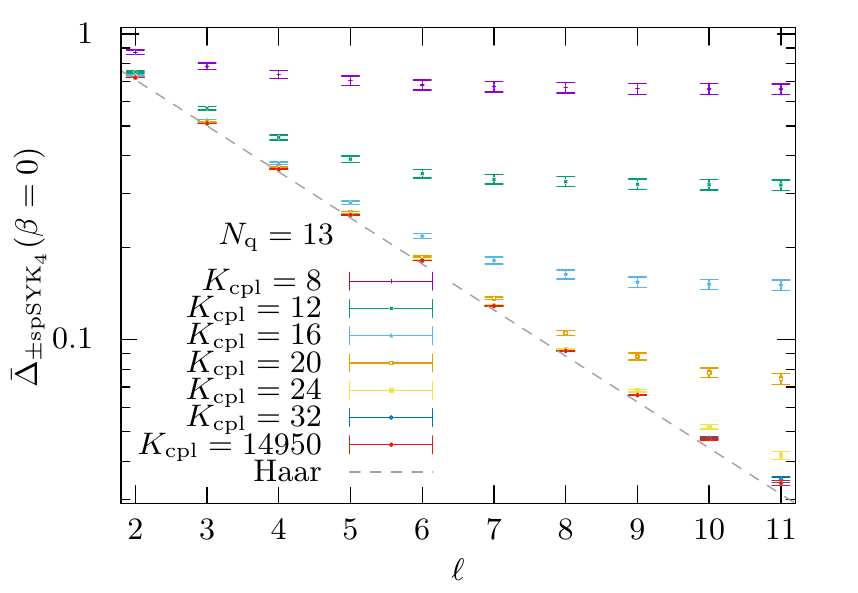}
    \caption{
    The late-time value of $\bar{\Delta}_{\pm \mathrm{spSYK}_4}(t,\beta=0)$ against $\ell$ is plotted for various numbers $\Kcpl$ of non-zero coupling constant. We set $\ND$ to $13$, and the number of samples is 64. The average for $t=(1,2,\ldots,10)\times 10^6$ is plotted as the late-time value in all figures.}
    \label{fig:Sparsek1lmax}
\end{figure}

In Fig.~\ref{fig:Sparsek1lmax}, we plot the upper bound on the recovery error for a further simplified sparse SYK model ($\pm \mathrm{spSYK}_4$), in which a half of the non-zero couplings is set to $1/\sqrt{\Kcpl}$ and the other half to $-1/\sqrt{\Kcpl}$~\cite{SparsePM1SYK}. We observe that, when $\Kcpl\gtrsim 30 = \cO(\ND)$,  this simplification does not change the upper bound on the recovery error from that of the Haar value.
Hence, $\pm \mathrm{spSYK}_4$ with $\Kcpl = \cO(\ND)$ suffices to reproduce information-theoretic properties of $\mathrm{SYK}_4$ as well as its chaotic features.
As this number of non-zero couplings is substantially smaller than the original $\mathrm{SYK}_4$, which has $\Kcpl = \cO(\ND^4)$, this would help experimental realizations of the Hayden-Preskill protocol in many-body systems.
See \ref{S:sparseSYK} \cite{Supplemental} for details.

\emph{Probing transitions by the Hayden-Preskill protocol.}
Yet another SYK model attracting much attention is the $\mathrm{SYK}_{4+2}$ model~\cite{GarciaGarcia2017PhysRevLett.120.241603}. The Hamiltonian is
\begin{equation}
    \hat{H}_{\mathrm{SYK}_{4+2}}(\theta) = \cos\theta \, \hat{H}_{\mathrm{SYK}_4}
    +\sin\theta \, \hat{H}_{\mathrm{SYK}_2},\label{eqn:SYK4+2}
\end{equation}
where $\hat{H}_{\mathrm{SYK}_2}=i \sum_{1\leq b_1<b_2\leq2\ND} K_{b_1b_2}\hat{\psi}_{b_1}\hat{\psi}_{b_2}$, and $\theta\in[0,\pi/2]$ is a mixing parameter.
The coupling constants $\{K_{b_1b_2}\}$ satisfy $K_{b_2b_1}=-K_{b_1b_2}$ and are normalized for the variance of eigenenergies of $\hat{H}_{\mathrm{SYK}_{4+2}}(\theta)$ to be unity.

The $\mathrm{SYK}_{4+2}$ model has a peculiar energy-shell structure in the sense of the local density of states in Fock space, which shows drastic changes by varying $\theta$. Accordingly, the range of $\theta \in [0, \pi/2]$ is divided into four regimes I, II, III, and IV~\cite{Monteiro2021PRR,Monteiro2021PRL}. In I, only one energy-shell is dominant in the whole Hilbert space, and it is quantum chaotic. As $\theta$ increases the size of the energy-shell becomes diminished, and $\cO({\rm poly} (\ND))$ energy-shells appear in II and III. The energy statistics remains RMT-like in these two regimes. Characterizing physics in II and III has been under intense investigations~\cite{NCDAR2022}. In IV, the number of energy-shell approaches $\cO(\exp (\ND))$, and Fock-space localization is observed.

\begin{figure}
    \centering
\includegraphics{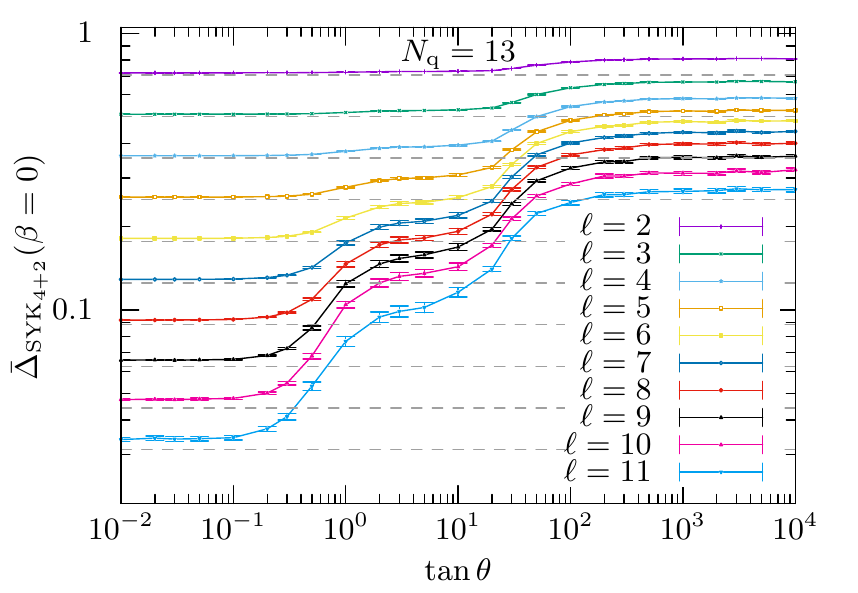}        
\caption{
    The late-time value of $\bar{\Delta}_{\mathrm{SYK}_{4+2}}(t,\beta=0)$ plotted for various $\ell$ against the value of $\delta$. $\ND$ is set to $13$. The number of samplings is 64. The lines connecting the data points are guide to the eye. The horizontal dashed lines indicate $\bar{\Delta}'_{\rm Haar}$ for various $\ell$.}
    \label{fig:delta_dependence}
\end{figure}

Based on the Hayden-Preskill protocol in $\mathrm{SYK}_{4+2}$, each regime can be operationally characterized.
In Fig.~\ref{fig:delta_dependence}, we plot the late time values of the upper bound $\bar{\Delta}_{\mathrm{SYK}_{4+2}}(t, \beta=0)$ of the recovery error against $\tan\theta$, in which two characteristic values of $\theta$, $\tan\theta_1 \approx 0.5$ and $\tan\theta_2 \approx 20$, are observed.
The first plateau ($\theta \in [0, \theta_1)$) corresponds to the regime I. As $\bar{\Delta}_{\mathrm{SYK}_{4+2}} \approx \bar{\Delta}'_{\rm Haar}$ in this regime, the system is information scrambling. 
The second one ($\theta \in (\theta_1, \theta_2]$) correspond to II and III, where $\bar{\Delta}_{\mathrm{SYK}_{4+2}}$ is substantially larger than $\bar{\Delta}'_{{\rm Haar}}$, which seems to be the case even in the large $\ND$ limit (see \ref{S:SYK4+2} \cite{Supplemental}). 
The third plateau, $\theta \in (\theta_2, \pi/2]$, corresponds to IV, where the system is almost $\mathrm{SYK}_{2}$. 

For sufficiently small and large $\theta$, the behaviour of $\bar{\Delta}_{\mathrm{SYK}_{4+2}}$ can be naturally understood. For small $\theta$, the model is approximately $\mathrm{SYK}_{4}$. As we have observed above, the dynamics of $\mathrm{SYK}_{4}$ quickly achieves the Hayden-Preskill recovery. Hence, this should also be the case in the regime I. In contrast, for sufficiently large $\theta$, the model is almost $\mathrm{SYK}_{2}$ and the Fock-space localization occurs. Thus, information recovery should not be possible, resulting in the absence of information scrambling in the regime IV.
In contrast, $\bar{\Delta}_{\mathrm{SYK}_{4+2}}$ is smoothly changing for the intermediate values of $\theta$, which is seemingly in tension with the division of the regimes II and III in terms of the energy-shell structure.

To understand the intermediate plateau, we shall recall that, in II and III, transitions from one energy-shell to the other are strongly suppressed, which effectively results in the division of the whole Hilbert space into $\cO({\rm poly}(\ND))$ energy-shells~\cite{Monteiro2021PRR}. Additionally, it is known that the dynamics in each energy-shell seems to be approximately Haar random \emph{within} the subspace~\cite{Monteiro2021PRL}. The intermediate plateau of $\bar{\Delta}_{\mathrm{SYK}_{4+2}}$ can be explained from these common features in II and III. Since the whole Hilbert space is effectively divided into smaller ones, within which the dynamics remains still Haar random, the unitary dynamics in II and III induces \emph{partial decoupling}~\cite{WN2021} rather than decoupling. In this case, the recovery error is given in the form of $2^{\ell'_{\rm th} - \ell} + \Delta_{\rm rem}$~\cite{NWK2022}. Here, $\ell'_{\rm th} \approx \ell_{\rm Haar, th} + \cO(\sqrt{k})$ and $\Delta_{\rm rem}$ quantifies the amount of information that cannot be recovered unless $\ell \approx \ND$. 
As we set $k=1$ in our numerics, $\ell'_{\rm th}$ is hardly observed in our analysis. In contrast, $\Delta_{\rm rem}$ is clearly observed as an intermediate plateau. 
It is known that $\Delta_{\rm rem}$ is inverse proportional to the standard deviation of energy in $D$. As the standard deviation of energy in $B$ shall be $\cO(\sqrt{\ND})$, that in $D$ is at least $\cO(\sqrt{\ND})$. Hence, we can qualitatively estimate that $\Delta_{\rm rem}=O(1/\sqrt{\ND})$, which remains nearly constant unless $\ell \approx \ND$. 

From this perspective, we can understand the two transitions as reflections of the changes of decoupling properties. In I, the combined regime over II and III, and IV, the $\mathrm{SYK}_{4+2}$ dynamics leads to full, partial, and no decoupling, respectively. Accordingly, each regime has qualitatively different behaviours in the information recovery.
The emerging difference between II and III in the energy-shell picture should be an artifact due to the fact that the energy-shell is viewed in the Fock basis, which is not necessarily physically intrinsic to the system.

\emph{Summary and discussions.}
In this Letter, we have studied the information recovery in various Hamiltonian systems and have shown that information scrambling in the sense of information recovery does not always coincide with quantum chaos. Spin chains are unlikely to be information scrambling, while they are quantum chaotic in energy spectrum and saturate OTOCs for local observables. In contrast, the (sparse) SYK models are information scrambling and have the latter two properties.
We have also demonstrated a potential use of the information recovery protocol to find new transitions caused by a information-theoretic mechanics. 

It is open if any local spin models can be information scrambling since the family of SYK models, the only models that are information scrambling in our analysis, does not have spatially local interactions.
It will be also of interest to further explore the direction of characterizing various quantum phases in the information-theoretic manner.\\

The work was supported by Grants-in-Aid for Transformative Research Areas (A) No.~JP21H05182, No.~JP21H05183 and No.~JP21H05185 from MEXT of Japan. The work of M.T. was partly supported by Grants-in-Aid No.~JP20H05270 and No.~JP20K03787 from MEXT of Japan. Y.N. was supported by JST, PRESTO Grant Number JPMJPR1865, Japan, and by JSPS KAKENHI Grant Number JP22K03464, Japan.
Part of the computation in this paper was conducted using the Supercomputing Facilities of the Institute for Solid State Physics, University of Tokyo.
The authors thank Satyam S. Jha for collaboration in the early stage of the work.
M.T. also thanks Masanori Hanada, Chisa Hotta, and Norihiro Iizuka for valuable discussions.

\bibliography{HPbySYK}

\newpage\hbox{}\thispagestyle{empty}\newpage
\onecolumngrid
\begin{center}
\textbf{\large Supplemental Materials: Hayden-Preskill Recovery in Hamiltonian Systems}
\end{center}
\setcounter{section}{0}
\setcounter{equation}{0}
\setcounter{figure}{0}
\setcounter{table}{0}
\setcounter{page}{1}
\makeatletter
\renewcommand{\thesection}{S\arabic{section}}
\renewcommand{\theequation}{S\arabic{equation}}
\renewcommand{\thefigure}{S\arabic{figure}}
\makeatother

\section{General analysis on the recovery error}\label{S:bounds}

We here provide a general analysis on the recovery error in the Hayden-Preskill protocol.  In~\ref{SS:ULbyDecoupling}, we provide upper and lower bounds on the recovery error based on the decoupling approach. 
The lower bound is, in general, computationally intractable. Hence, we derive another lower bound based on the mutual information in~\ref{SS:LBbyMI}.

\subsection{Upper and lower bounds by decoupling} \label{SS:ULbyDecoupling}

A decoupling condition provides necessary and sufficient conditions for the recovery of quantum information~\cite{doi:10.1142/S1230161208000043,dupuis_one-shot_2014,Dupuis1004.1641}.
We here briefly explain how upper and lower bounds on the recovery error are obtained from the decoupling approach.

When the system Hamiltonian in $S$ is $\hat{H}_S$, the state at time $t$ in the Hayden-Preskill protocol is
\begin{equation}
    \ket{\Psi(t, \beta)}^{S B' R}
    =
    e^{-i \hat{H}_S t} \bigl(\ket{\Phi}^{AR} \otimes \ket{\xi(\beta)}^{B B'} \bigr).
\label{eqn:Psifin}
\end{equation}
Here, $\ket{\xi(\beta)}^{B B'}$ is a pure state, whose reduced state in $B$ is a thermal state $\xi^B(\beta)$ at the inverse temperature $\beta$.
The recovery error of the Hayden-Preskill protocol is defined as
\begin{equation}
    \Delta_{\hat{H}}(t, \beta) = \frac{1}{2} \min_{\cD} 
    \norm{
    \Phi^{AR} - \cD^{C B' \rightarrow A} \bigl( \Psi^{C B' R}(t) \bigr) 
    }_1,
\end{equation}
where the minimum is taken over all completely-positive and trace-preserving (CPTP) maps $\cD$ from $C B'$ to $A$, $\Phi^{AR} = \ketbra{\Phi}{\Phi}^{AR}$ is the maximally entangled state between $A$ and $R$, and $\Psi^{C B' R}(t, \beta) = \Tr_{D} [\Psi^{S B' R}(t, \beta)]$. Note that the system $S$ can be decomposed into $AB$ as well as $CD$.

We below show
\begin{equation}
    1- \sqrt{1-\bigl(\Sigma^{\rm opt}_{\hat{H}}(t, \beta)\bigr)^2}
    \leq
    \Delta_{\hat{H}}(t, \beta)
    \leq 
    \min\bigl\{ 1,\sqrt{2\Sigma^{\rm opt}_{\hat{H}}(t, \beta)} \bigr\}. \label{Eq:UpperBound0}
\end{equation}
Here, $\Sigma^{\rm opt}_{\hat{H}}(t, \beta)$ is the degree of decoupling between $D$ and $R$, defined by
\begin{equation}
    \Sigma^{\rm opt}_{\hat{H}}(t, \beta) :=\frac12
    \min_{\sigma}
    \norm{ \Psi^{D R}(t, \beta)
    -
    \sigma^{D} \otimes \pi^R
    }_1, \label{Eq:eerrrtf}
\end{equation}
where the minimization is taken over all quantum states on $D$, and $\Psi^{D R}(t, \beta) = \Tr_{C} [ \Psi^{C D R}(t, \beta) ]$.
We observe from Eq.~\eqref{Eq:UpperBound0} that the recovery error $\Delta_{\hat{H}}(t, \beta)$ is small if and only if $\Sigma^{\rm opt}_{\hat{H}}(t, \beta) \ll 1$.
In the derivation below, we do not explicitly write $(t, \beta)$. 

Let $T(\rho, \sigma)$ be the trace distance between $\rho$ and $\sigma$, i.e., $T(\rho, \sigma) = \frac{1}{2}\| \rho - \sigma\|_1$. For pure states $\ket{\phi}$ and $\ket{\psi}$, we denote the trace distance by $T(\ket{\phi}, \ket{\psi}) = \frac{1}{2} \norm{ \ketbra{\phi}{\phi}- \ketbra{\psi}{\psi} }_1$ for simplicity.
To obtain the bounds in Eq.~\eqref{Eq:UpperBound0}, we use the relation between the trace distance and the fidelity:
\begin{equation}
    1- \sqrt{F(\rho, \sigma)} \leq T(\rho, \sigma) \leq \sqrt{1- F(\rho, \sigma)}, \label{Eq:TDF}
\end{equation}
where $F(\rho, \sigma) : = \| \sqrt{\rho} \sqrt{\sigma} \|_1^2$ is the fidelity.
We also use the Uhlmann's theorem:
\begin{equation}
    F(\rho^A, \sigma^A) = \max_V \bigl|\bra{\rho}^{AB} V^{C \rightarrow B} \ket{\sigma}^{AC} \bigr|^2, \label{Eq:Uhlmann}
\end{equation}
where the maximization is over all isometries $V^{C \rightarrow B}$ from $C$ to $B$ ($\dim \mathcal{H}_C \leq \dim \mathcal{H}^B$), and $\ket{\rho}^{AB}$ and $\ket{\sigma}^{AC}$ are purifications of $\rho^A$ and $\sigma^A$, respectively.

The upper bound in Eq.~\eqref{Eq:UpperBound0} directly follows from the Uhlmann's theorem in terms of the trace norm (see, e.g., \cite{Dupuis1004.1641}). It states that, if there are two states $\rho^A$ and $\sigma^A$ such that $T(\rho, \sigma) \leq \epsilon$, then there exists systems $B$ ($C$) that purifies $\rho^A$ ($\sigma^A$) to a pure state $\ket{\rho}^{AB}$ ($\ket{\sigma}^{AC})$ and a partial isometry $V^{B \rightarrow C}$ that satisfies $T(V^{B \rightarrow C} \ket{\rho}^{AB}, \ket{\sigma}^{AC}) \leq \sqrt{2 \epsilon}$.
We apply this relation to Eq.~\eqref{Eq:eerrrtf}.
A purification of $\Psi^{D R}$ and $\pi^R$ are $\ket{\Psi}^{S B' R}$ and $\ket{\Phi}^{\hat{A}R}$, respectively.
We denote by $\ket{\sigma}^{D E}$ a purification of $\sigma^{D}$ in Eq.~\eqref{Eq:eerrrtf} by some system $E$. Then, there is a partial isometry $V^{D B' \rightarrow \hat{A}E}$ such that
\begin{equation}
    T \bigl( V^{CB' \rightarrow \hat{A}E} \ket{\Psi}^{S B' R}, \ket{\sigma}^{D E} \otimes \ket{\Phi}^{\hat{A}R}  \bigr) \leq \sqrt{2 \Sigma^{\rm opt}_{\hat{H}}}.
\end{equation}
By tracing out $D E$, we obtain
\begin{equation}
    T \bigl( \cD^{C B' \rightarrow \hat{A}} (\Psi^{C B' R}),  \Phi^{\hat{A}R}  \bigr) \leq \sqrt{2 \Sigma^{\rm opt}_{\hat{H}}},
\end{equation}
where $\cD$ is a CPTP map obtained from the isometry $V$ and the partial trace.
As the left-hand side is exactly $\Delta_{\hat{H}}$ by the identification of the system $\hat{A}$ with $A$, this provides an upper bound on the recovery error.

The lower bound in Eq.~\eqref{Eq:UpperBound0} is obtained using Eqs.~\eqref{Eq:TDF} and~\eqref{Eq:Uhlmann}:
\begin{align}
    \Delta_{\hat{H}}
    &:=\min_{\cD} T \bigl(\Phi^{AR}, \cD^{C B' \rightarrow A} \bigl( \Psi^{C B' R} \bigr) \bigr)\\
    &\geq \min_{\cD} \bigl\{ 1 - \sqrt{F\bigl(\Phi^{AR}, \cD^{C B' \rightarrow A} ( \Psi^{C B' R})} \bigr) \bigr\}\\
    &= \min_{V, \sigma} \bigl\{ 1 - |\bra{\Psi}^{S B' R}  V^{AS' \rightarrow B' C}(\ket{\Phi}^{AR} \otimes \ket{\sigma}^{DS'}) | \bigr\}\\
    & \geq \min_{\sigma} \bigl\{ 1 - \sqrt{F(\Psi^{D R}, \pi^R \otimes \sigma^{D})} \bigr\}\\
    & \geq \min_{\sigma} \bigl\{ 1 - \sqrt{ 1 - \bigl(T(\Psi^{D R}, \pi^R \otimes \sigma^{D}) \bigr)^2} \bigr\}\\
    & \geq 1 - \sqrt{1 - \bigl(\Sigma^{\rm opt}_{\hat{H}} \bigr)^2}.
\end{align}
Here, the second and the second last lines follow from Eq.~\eqref{Eq:TDF}, the third line from the Uhlmann's theorem, and the fourth line from the monotonicity of the fidelity under the partial trace.

While the bounds in Eq.~\eqref{Eq:UpperBound0} replace the analysis of the recovery error with that of decoupling, the degree of decoupling $\Sigma^{\rm opt}_{\hat{H}}$ is, in general, computationally intractable due to the minimization over all states in the system $D$. 
To circumvent this issue, the state $\sigma$ to be minimized in Eq.~\eqref{Eq:eerrrtf} is commonly set to $\Psi^{D}$. By doing so, we obtain a computationally tractable upper bound as
\begin{equation}
    \Delta_{\hat{H}}
    \leq 
    \bar{\Delta}_{\hat{H}}
    := 
    \min \bigl\{1, \sqrt{2\Sigma_{\hat{H}}} \bigr\}, \label{Eq:UpperBoundrrrr}
\end{equation}
where $\Sigma_{\hat{H}} :=  \frac{1}{2}\norm{\Psi^{D R} - \Psi^{D} \otimes \pi^R}_1$. This is the bound we have used throughout the paper. Note that a computationally tractable lower bound cannot be obtained in this approach.

\subsection{Upper and lower bounds from the mutual information} \label{SS:LBbyMI}
Instead of the decoupling approach, we may use the mutual information of the state $\ket{\Psi(t, \beta)}^{SB'R}$ to bound the recovery error, leading to
\begin{equation}
    1- \sqrt{1- \bigl(f^{-1}\bigl( I(D:R)_{\Psi} \bigr) \bigr)^2} \leq \Delta_{\hat{H}} \leq \min\{1, \bigl(2 \ln 2 I(D:R)_{\Psi} \bigr)^{1/4}\}.\label{Eq:ULboundsMI}
\end{equation}
Here, $I(D:R)_{\Psi} = H(R)_{\Psi} - H(R|D)_{\Psi}$, where $H(R)_{\Psi} = {\rm Tr}[-\Psi^R \log \Psi^R]$ is the von Neumann entropy of the state $\Psi^R$, and $H(R|D)_{\Psi} = H(DR)_{\Psi} - H(D)_{\Psi}$ is the conditional entropy. For simplicity, we omitted $(t, \beta)$.
Note that the mutual information $I(D:R)_{\Psi}$ satisfies $I(D:R)_{\Psi} +  I(CB':R)_{\Psi} = 2k$ since $\ket{\Psi}^{DCB'R}$ is pure. Hence, the mutual information $I(D:R)_{\Psi}$ in the bounds (Eq.~\eqref{Eq:ULboundsMI}) can be replaced with $2k - I(R:CB')_{\Psi}$.

To derive the upper bound in Eq.~\eqref{Eq:ULboundsMI}, we simply use the relation between the quantity $\Sigma_{\hat{H}}$, defined just below Eq.~\eqref{Eq:UpperBoundrrrr}, and the mutual information $I(D:R)_{\Psi}$, namely, $\Sigma_{\hat{H}} \leq \sqrt{\frac{\ln 2}{2} I(D:R)_{\Psi}}$ (see, e.g., \cite{wilde_2013}). From this and the upper bound on the recovery error in Eq.~\eqref{Eq:UpperBoundrrrr}, we obtain
\begin{equation}
    \Delta_{\hat{H}} \leq \min\{1, \bigl(2 \ln 2 I(D:R)_{\Psi} \bigr)^{1/4}\}.
\end{equation}

As for the lower bound, let $\hat{\sigma}^D$ be ${\rm argmin} \frac{1}{2}\| \Psi^{D R} - \sigma^{D} \otimes \pi^R \|_1$, and $\hat{\Psi}^{DR}$ be $\hat{\sigma}^D \otimes \pi^R$. Using the fact that $I(D:R)_{\hat{\Psi}} = 0$, the mutual information can be rewritten as 
\begin{align}
    I(D:R)_{\Psi} &= I(D:R)_{\Psi} - I(D:R)_{\hat{\Psi}}\\
    &= H(R)_{\Psi} - H(R|D)_{\Psi} - \bigl(H(R)_{\hat{\Psi}} - H(R|D)_{\hat{\Psi}}\bigr)\\
    &= H(R|D)_{\hat{\Psi}} - H(R|D)_{\Psi},
\end{align}
where we used that $H(R)_{\Psi} = H(R)_{\hat{\Psi}}$ as $\Psi^R = \hat{\Psi}^R = \pi^R$.
Applying the Alicki-Fannes-Winter inequality~\cite{RAlicki_2004,Winter2016}, the last expression is bounded from above by a function of the trace norm between $\Psi^{DR}$ and $\hat{\Psi}^{DR}$, which is nothing but $\Sigma^{\rm opt}_{\hat{H}}$. We then have $I(D:R)_{\Psi} \leq f(\Sigma^{\rm opt}_{\hat{H}})$, where $f(x) = 2 k x  + (1+x)h\bigl( \frac{x}{1+x} \bigr)$, and $h(x) = -x \log x - (1-x) \log (1-x)$ for $0\leq x \leq 1$ is the binary entropy. 
Since $f(x)$ is monotonically increasing function for any $x \geq 0$, it has the inverse $f^{-1}$, and we have $   f^{-1}\bigl( I(D:R)_{\Psi} \bigr) \leq \Sigma^{\rm opt}_{\hat{H}}$. Together with the lower bound on $\Delta_{\hat{H}}(t, \beta)$ in terms of $\Sigma^{\rm opt}_{\hat{H}}$~\eqref{Eq:UpperBound0}, we have
\begin{equation}
    1- \sqrt{1- \bigl(f^{-1}\bigl( I(D:R)_{\Psi} \bigr) \bigr)^2} \leq \Delta_{\hat{H}}. \label{Eq:lowerrecoveryerror}
\end{equation}

It is obvious that Eq.~\eqref{Eq:ULboundsMI} is in general worse than Eq.~\eqref{Eq:UpperBound0} because the former is obtained by substituting into Eq.~\eqref{Eq:UpperBound0} the bounds on the degree of decoupling, $\Sigma_{\hat{H}}^{\rm opt}$, in terms of the mutual information $I(D:R)_{\Psi}$. However, the lower bound in Eq.~\eqref{Eq:ULboundsMI} has an advantage from a computational viewpoint since it is calculable if the mutual information is.
Based on this consideration, we used in the main text an upper and a lower bound on the recovery error such as
\begin{equation}
    1- \sqrt{1- \bigl(f^{-1}\bigl( I(D:R)_{\Psi} \bigr) \bigr)^2} =  \underline{\Delta}_{\hat{H}} \leq \Delta_{\hat{H}} \leq \bar{\Delta}_{\hat{H}}
    = 
    \min \bigl\{1, \sqrt{2\Sigma_{\hat{H}}} \bigr\}. \label{Eq:KoregaWarewareNoULbounds}
\end{equation}

Before we leave this section, we point out that $I(D:R)_{\Psi}$ has been studied as the OMI, especially for the infinite temperature ($\beta =0$)~\cite{Hosur_2016,caputa_scrambling_2015,Nie_2019,https://doi.org/10.48550/arxiv.2302.08009}. 
Using Eq.~\eqref{Eq:ULboundsMI}, the results in the literature can be rephrased in terms of the recovery error.

\section{Commuting Hamiltonian models} \label{S:CH}
We provide an in-depth analysis of the Hayden-Preskill protocol for  commuting Hamiltonians. We mean by a commuting Hamiltonian the one in the form of
\begin{equation}
    \hat{H}_S = \hat{H}_A + \hat{H}_{B} + \hat{H}_{A:B}, \label{Eq:decomp}
\end{equation}
where $\hat{H}_A$ and $\hat{H}_{B}$ are Hamiltonians non-trivially acting only on $A$ and $B$, respectively, and $\hat{H}_{A:B}$ is an interacting Hamiltonian between $A$ and $B$ that commutes with the other two. Below, we assume that $A \subseteq D$, which implies $C \subseteq B$ as $S=AB=CD$.

Due to the commuting condition, the time-evolving operator generated by $\hat{H}_S$ is decomposed to $e^{-i \hat{H}_S t} =e^{-i \hat{H}_{A:B} t} e^{-i \hat{H}_A t} e^{-i \hat{H}_{B} t} $.
Since a thermal state $\xi(\beta)^{B} \propto e^{- \beta \hat{H}_{B}}$ is invariant under the time-evolution by $\hat{H}_{B}$, we have $\Psi^{D R}(t, \beta) = \Tr_{C} \bigl[ e^{-i \hat{H}_{A:B} t} e^{-i \hat{H}_A t} \bigl(\Phi^{AR} \otimes \xi(\beta)^{B} \bigr)   e^{i \hat{H}_A t}e^{i \hat{H}_{A:B} t} \bigr]$.
We then use the property of maximally entangled state that, for any unitary $U^A$ on the system $A$, $U^A \ket{\Phi}^{AR} = \bar{U}^R \ket{\Phi}^{AR}$, where $\bar{\dot}$ indicates the complex conjugate. Denoting $e^{-i \hat{H}_A t}$ by $U^A(t)$, it follows that $\Psi^{D R}(t, \beta) = \bar{U}^R \Tr_{C} \bigl[ e^{-i \hat{H}_{A:B} t} \bigl(\Phi^{AR} \otimes \xi(\beta)^{B} \bigr) e^{i \hat{H}_{A:B} t} \bigr]  (\bar{U}^R)^{\dagger}$.
This leads to
\begin{align}
    \Sigma^{\rm opt}_{\hat{H}}(t, \beta) 
    &=
    \frac{1}{2}\min_{\sigma} 
    \norm{ \bar{U}^R \Tr_{C} \bigl[ e^{-i \hat{H}_{A:B} t} \bigl(\Phi^{AR} \otimes \xi(\beta)^{B} \bigr) e^{i \hat{H}_{A:B} t} \bigr]  (\bar{U}^R)^{\dagger} - \sigma^{D} \otimes \pi^R }_1,\\
    &=
    \frac{1}{2}\min_{\sigma} 
    \norm{ \Tr_{C} \bigl[ e^{-i \hat{H}_{A:B} t} \bigl(\Phi^{AR} \otimes \xi(\beta)^{B} \bigr) e^{i \hat{H}_{A:B} t} \bigr] - \sigma^{D} \otimes \pi^R }_1,\label{Eq:aarrccerr}
\end{align}
where we used the unitary invariance of the trace norm and $\pi^R = I^R/2^k$ in the last line.

We now introduce an extended region $\bar{B}$ of $B$ in terms of $\hat{H}_{A:B}$. Namely, $\bar{B}$ is the union of the system $B$ and the set of qubits on which $\hat{H}_{A:B}$ acts non-trivially. As we assumed that $C \subseteq B$, it holds that $C \subseteq \bar{B}$.
We denote by $\partial \bar{B}$ the boundary of $B$ in terms of $\hat{H}_{A:B}$, i.e., $\partial \bar{B}=\bar{B} \setminus B$.
By taking the trace over $\bar{B} \setminus C$ in the right-hand side Eq.~\eqref{Eq:aarrccerr} and using the monotonicity of the trace norm under the partial trace, we have\begin{align}
    \Sigma^{\rm opt}_{\hat{H}}(t, \beta)
    &\geq\frac{1}{2} \min_{\sigma} \norm{ \Tr_{\bar{B}} \bigl[e^{-i \hat{H}_{A:B} t} \bigl(\Phi^{AR} \otimes \xi(\beta)^{B} \bigr) e^{i \hat{H}_{A:B} t} \bigr] - \Tr_{D \cap \bar{B}} [\sigma^{D}] \otimes \pi^R }_1\nonumber\\
    &= \frac{1}{2}\min_{\sigma} 
    \norm{ 
    \Tr_{ \partial \bar{B}} \bigl[ \Phi^{AR} ] - \sigma^{A \setminus \partial \bar{B}} \otimes \pi^R 
    }_1,
\end{align}
where the last line holds since the interaction Hamiltonian $\hat{H}_{A:B}$ nontrivially acts only within $\bar{B}$.

Let $A_-$ be $A \setminus \partial \bar{B}$. As $\partial \bar{B} \subseteq A$, $A_-$ is not empty in general. Using this notation, we have $\Tr_{ \partial \bar{B}} \bigl[ \Phi^{AR} ] = \Phi^{A_- R_-} \otimes \pi^{R_+}$, where $R$ is divided into $R_-R_+$, $\Phi^{A_- R_-}$ is a maximally entangled state between $A_-$ and $R_-$, and $\pi^{R_+}$ is the completely mixed state in $R_+$. Using $\kappa$, defined by the number of qubits in $A$ that interact with $B$ by $\hat{H}_{A:B}$, $\Phi^{A_- R_-}$ is simply $(k-\kappa)$ EPR pairs.
As $\pi^R = \pi^{R_-} \otimes \pi^{R_+}$, it follows that
\begin{align}
    \Sigma^{\rm opt}_{\hat{H}}(t, \beta)
    &\geq \frac{1}{2}\min_{\sigma} 
    \norm{ 
    (\Phi^{A_- R_-} - \sigma^{A_-} \otimes \pi^{R_-}) \otimes \pi^{R_+} 
    }_1,\nonumber\\
    &= \frac{1}{2}\min_{\sigma} 
    \norm{ 
    \Phi^{A_- R_-} - \sigma^{A_-} \otimes \pi^{R_-}
    }_1.
\end{align}
Using Eq.~\eqref{Eq:TDF} and the fact that $F(\Phi^{A_- R_-}, \sigma^{A_-} \otimes \pi^{R_-}) = 1/2^{2(k-\kappa)}$, we obtain $\Sigma^{\rm opt}_{\hat{H}}(t, \beta) \geq 1 -  2^{\kappa - k}$. Substituting this into the lower bound in Eq.~\eqref{Eq:UpperBound0}, we have
\begin{equation}
    \Delta_{\hat{H}} \geq 1 - \sqrt{2^{\kappa - k}(2-2^{\kappa - k})}. \label{Eq:a;oierw}
\end{equation}
Thus, when $\kappa < k$, the recovery error $\Delta_{\hat{H}}$ is bounded from below by a constant.

Although the lower bound in Eq.~\eqref{Eq:a;oierw} becomes trivial when $\kappa = k$, this is just due to the analytical derivation. To illustrate this, let us particularly consider the Sherrington-Kirkpatrick (SK) Hamiltonian given by
$\hat{H}_{\rm SK} := - \sum_{n < m} J_{nm} Z_n \otimes Z_m$, where $J_{nm}$ is chosen from a given distribution. 
Similarly to Eq.~\eqref{Eq:decomp}, we first divide the Hamiltonian into three, such as $\hat{H}_{\rm SK} = \hat{H}_A + \hat{H}_{B} + \hat{H}_{A:B}$, where $\hat{H}_A$ and $\hat{H}_{B}$ non-trivially act only on $A$ and $B$, respectively.
Following the above argument, it holds that
\begin{align}
    \Sigma^{\rm opt}_{\hat{H}_{\rm SK}}(t, \beta)
    &=\frac{1}{2}\min_{\sigma} \norm{ \Tr_{C} \bigl[ e^{-i \hat{H}_{A:B} t} (\Phi^{AR} \otimes \xi(\beta)^{B}) e^{i \hat{H}_{A:B} t} \bigr] - \sigma^{D} \otimes \pi^R }_1.
\end{align}
We further decompose $\hat{H}_{A:B}$ into $\hat{H}_{A:B} = \hat{H}_{(A:B) \cap D}+ \hat{H}_{A:C}$, where the former term non-trivially acts only on $D$ and the latter is the rest. The unitary invariance of the trace norm leads to
\begin{align}
    \Sigma^{\rm opt}_{\hat{H}_{\rm SK}}(t, \beta)
    &=\frac{1}{2}\min_{\sigma} \norm{ 
    \Tr_{C} \bigl[ e^{-i \hat{H}_{A:C} t} (\Phi^{AR} \otimes \xi(\beta)^{B}) e^{i \hat{H}_{A:C} t} \bigr]  - \sigma^{D} \otimes \pi^R }_1.
\end{align}

By applying a CPTP map onto $R$ that maps $\rho^R$ to $\sum_{j} \bra{j} \rho^R\ket{j} \ketbra{j}{j}^R$, where $\{ \ket{j} \}_{j}$ ($j = 0, \dots, 2^k-1$) is the Pauli-$Z$ basis in $R$, we have
\begin{align}
    \Sigma^{\rm opt}_{\hat{H}_{\rm SK}}(t, \beta)  
    &\frac{1}{2}\geq \min_{\sigma} \norm{ \Tr_{C} \bigl[ e^{-i \hat{H}_{A:C} t} (\Omega^{AR} \otimes \xi(\beta)^{B}) e^{i \hat{H}_{A:C} t}]  - \sigma^{D} \otimes \pi^R }_1\nonumber\\
    &\geq \frac{1}{2} \min_{\sigma} \norm{ \Tr_{B} \bigl[ e^{-i \hat{H}_{A:C} t} (\Omega^{AR} \otimes \xi(\beta)^{B}) e^{i \hat{H}_{A:C} t}] 
    - \sigma^{A} \otimes \pi^R }_1,
\end{align}
where $\Omega^{AR} := 2^{-k} \sum_{j} \ketbra{j}{j}^A \otimes \ketbra{j}{j}^R$. Note that $\pi^R$ is invariant under the above CPTP map, and that the second inequality follows as $C \subseteq B$.
We now use the relation that, for $n \in A$ and $m \in C$,
\begin{equation}
    e^{i J_{nm} Z_n \otimes Z_m t} (|j\rangle \langle j|^A \otimes \xi(\beta)^{B}) e^{-i J_{nm} Z_n \otimes Z_m t}
    =
    \ketbra{j}{j}^A \otimes e^{(-1)^{j_n} i J_{nm} Z_m} \xi(\beta)^{B} e^{-(-1)^{j_n} i J_{nm} Z_m},
\end{equation}
where we express $j$ in binary as $j_1 \dots j_k$ ($j_n = 0,1$). The terms such as $e^{(-1)^{j_n} i J_{nm} Z_m}$ then disappears when $\Tr_{B}$ is taken.
Hence, it follows that 
\begin{equation}
    \Tr_{B} \bigl[ e^{-i \hat{H}_{A:C} t} (\Omega^{AR} \otimes \xi(\beta)^{B}) e^{i \hat{H}_{A:C} t}]
    =
    \Omega^{AR}.
\end{equation}
Hence, we have
\begin{align}
    \Sigma^{\rm opt}_{\hat{H}_{\rm SK}}(t, \beta) 
    &\geq \frac{1}{2} \min_{\sigma^A} \norm{ \Omega^{AR}  - \sigma^A \otimes \pi^R }_1\\
    &\geq \min_{\sigma^A}  \bigl[1 - \sqrt{F(\Omega^{AR}, \sigma^A \otimes \pi^R)} \bigr]\\
    &\geq 1- 2^{-k/2},
\end{align}
where we have used the lower bound given in Eq.~\eqref{Eq:TDF} and the last line follows from the direct calculation. This leads to $\Delta_{\hat{H}_{\rm SK}} \geq 1- \sqrt{2^{-k/2}(2- 2^{-k/2})}$.

\section{The Heisenberg model with random magnetic field} \label{S:Heisenberg}
We consider a one-dimensional quantum spin chain with site-dependent random magnetic field to the $z$ direction. The Hamiltonian is
\begin{align}
    \hat{H}_\mathrm{XXZ}
    &= \frac14\sum_{j=1}^{N-1} \left(X_{j}X_{j+1}+Y_{j}Y_{j+1}+J_z Z_{j}Z_{j+1}\right)+\frac12\sum_{j=1}^N h_j Z_j,
\end{align}
in which $N$ is the number of $S=1/2$ spins, $h_j$ are independently sampled from a uniform distribution in $[-W,W]$, $J_z$ is the ratio of the coupling in the $z$ direction to that in the $xy$ plane.

The model has different features depending on the choice of $J_z$ and $W$. For $J_z=0$ and $h_j=0$ ($j=1,2,\ldots,N$), the Hamiltonian becomes
\begin{align}
    \hat{H}_\mathrm{XY}
    = \frac14\sum_{j=1}^{N-1} \left(X_{j}X_{j+1}+Y_{j}Y_{j+1}\right)
    = \sum_{j=1}^{N-1} \left(S_{j}^xS_{j+1}^x+S_{j}^yS_{j+1}^y\right),
\end{align}
which is integrable.
For $J_z=1$ and $W>0$, the model has been extensively studied as a prototypical model of MBL in one spatial dimension \cite{LFSantos_2004, Kudo_Deguchi_PhysRevB.69.132404,Viola_Brown_JPhysA_2007,Znidaric_Prosen_Prelovsek_PhysRevB.77.064426,Pal_Huse_PhysRevB.82.174411.2010,De_Luca_Scardicchio_EPL_2013,Alet_Laflorencie_2018,RevModPhys.91.021001}.
For system sizes accessible by numerical diagonalization, various measures of localization point to the MBL transition at finite critical $W$ \cite{Luitz-Laflorencie-Alet-PhysRevB.91.081103}.
Note, however, that the location of the transition to the genuine MBL phase in the thermodynamic limit have been heavily debated in more recent studies.\cite{Morningstar_Huse_PRB2019,Suntajs_Bonca_Prosen_Vidmar_PhysRevE.102.062144,Sierant_Delande_Zakrzewski_2020,Kiefer-Emmanouilidis-PhysRevLett.124.243601,Chanda_Sierant_Zakrzewski_PhysRevResearch.2.032045,Sels-Polkovnikov-PhysRevE.104.054105,Kiefer-Emmanouilidis-PhysRevB.103.024203,Morningstar__PhysRevB.105.174205,Sels_PhysRevB.106.L020202,Sierant_Zakrzewski_PhysRevB.105.224203,Ghosh_Znidaric_PRB2022} 

In the following, we compute the upper and lower bounds on the recovery error for this model based on Eq.~\eqref{Eq:KoregaWarewareNoULbounds}, which are denoted by $\bar{\Delta}_{\rm XXZ}$ and $\underbar{$\Delta$}_{\rm XXZ}$, respectively.
When $J_z=0$, we denote the bounds by $\bar{\Delta}_{\rm XY}$ and $\underbar{$\Delta$}_{\rm XY}$.

\begin{figure}
    \centering
    \includegraphics[width=\linewidth]{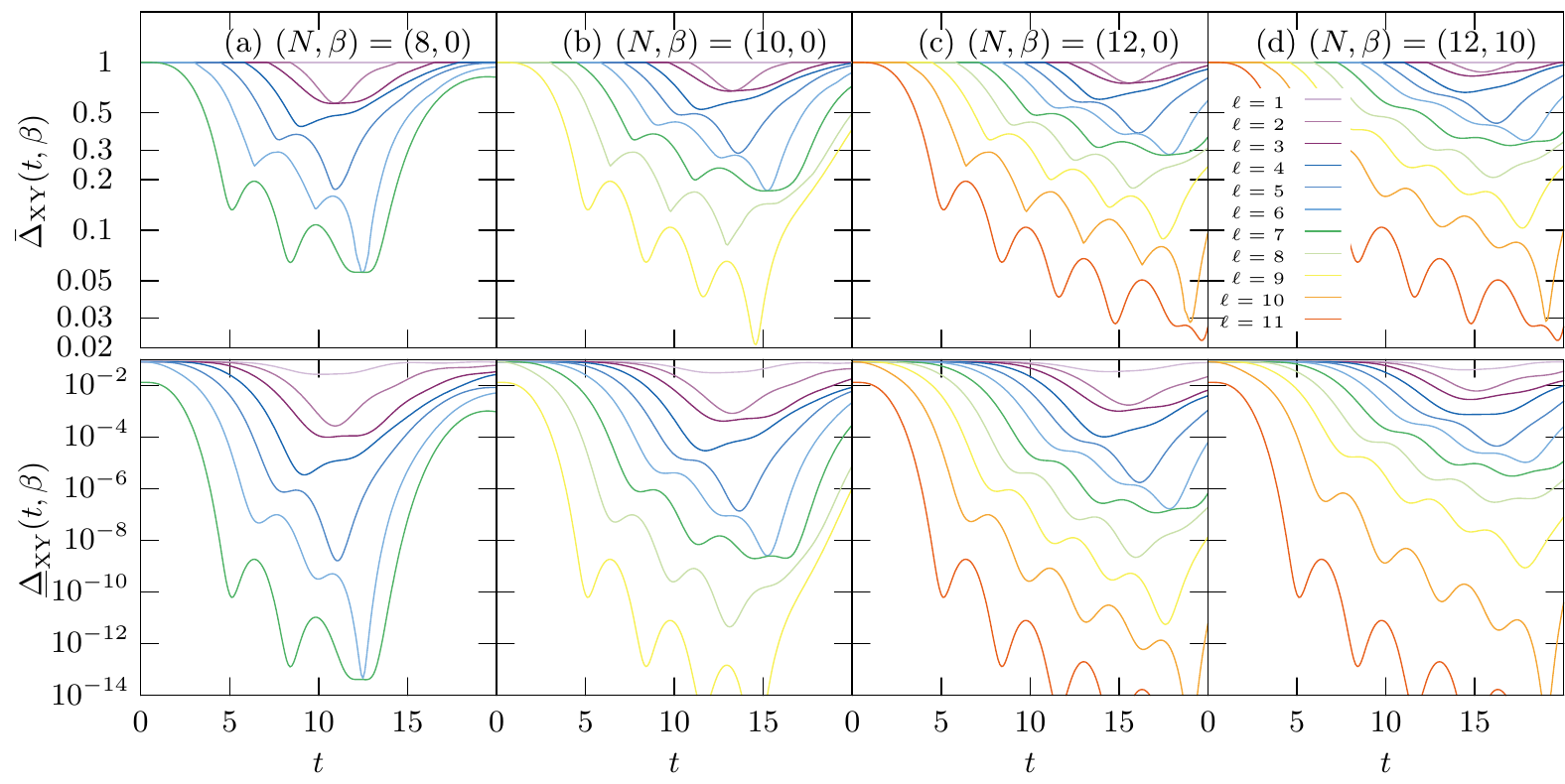}
    \caption{
    The values of $\bar{\Delta}_{\rm XY}, \underline{\Delta}_{\rm XY}$ for the XY model with 
    $(N,\beta)=$ (a) $(8,0)$, (b) $(10,0)$, (c) $(12,0)$, and (d) $(12,10)$.
    }
    \label{fig:XY_model}
\end{figure}

\begin{figure}
    \centering
    \includegraphics[width=\linewidth]{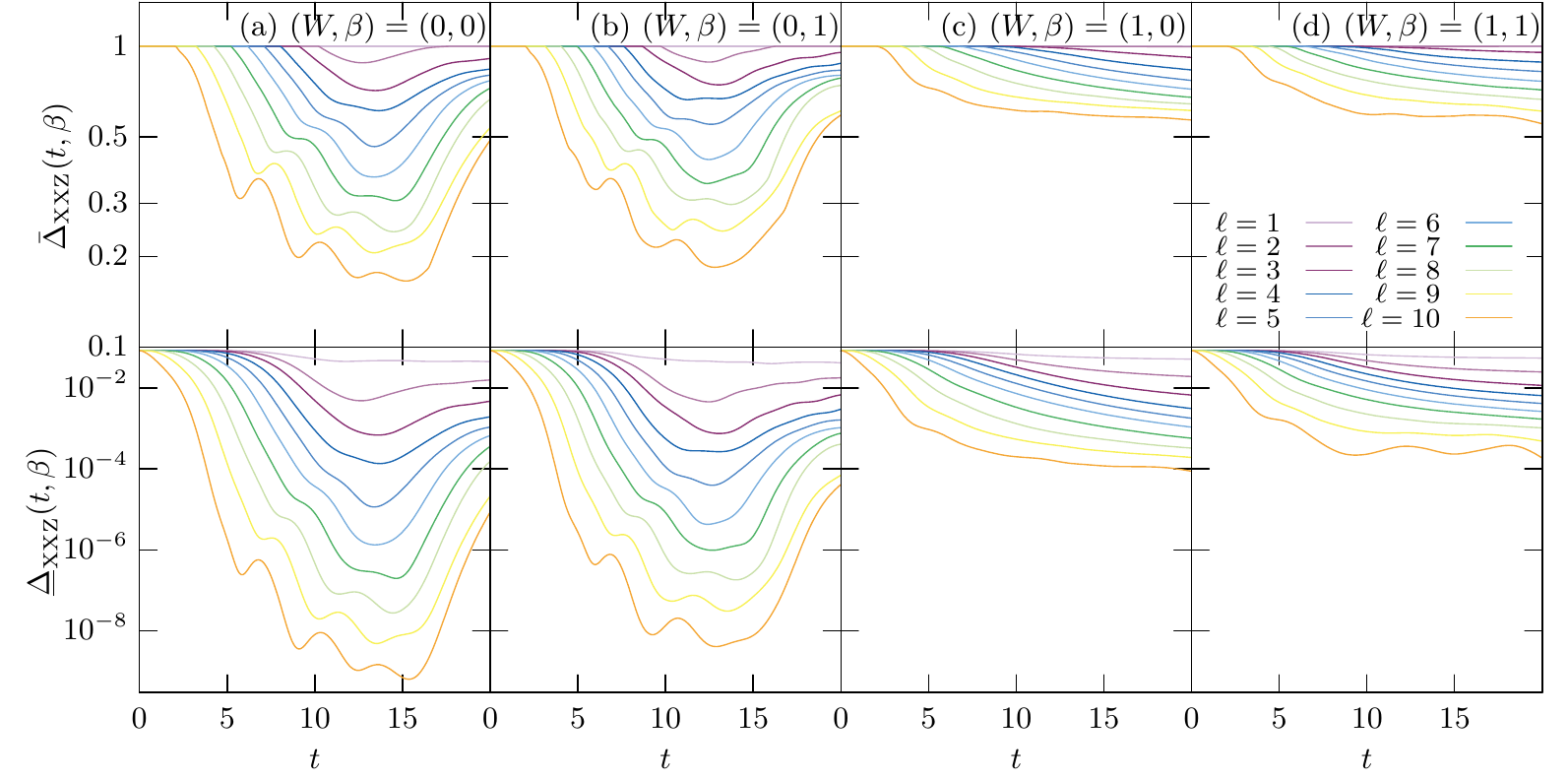}
    \caption{
    The values of $\bar{\Delta}_{\rm XXZ}$ and $\underline{\Delta}_{\rm XXZ}$  for the XXZ model with $J_z=1$, $N=12$ and
    $(W,\beta)=$ (a) $(0,0)$, (b) $(0,1)$, (c) $(1,0)$, and (d) $(1,1)$.
    For (c) and (d), averages over 16 samples are plotted for $W>0$.
    }
    \label{fig:XXZ_W0-1}
\end{figure}

\begin{figure}
    \centering
    \includegraphics[width=\linewidth]{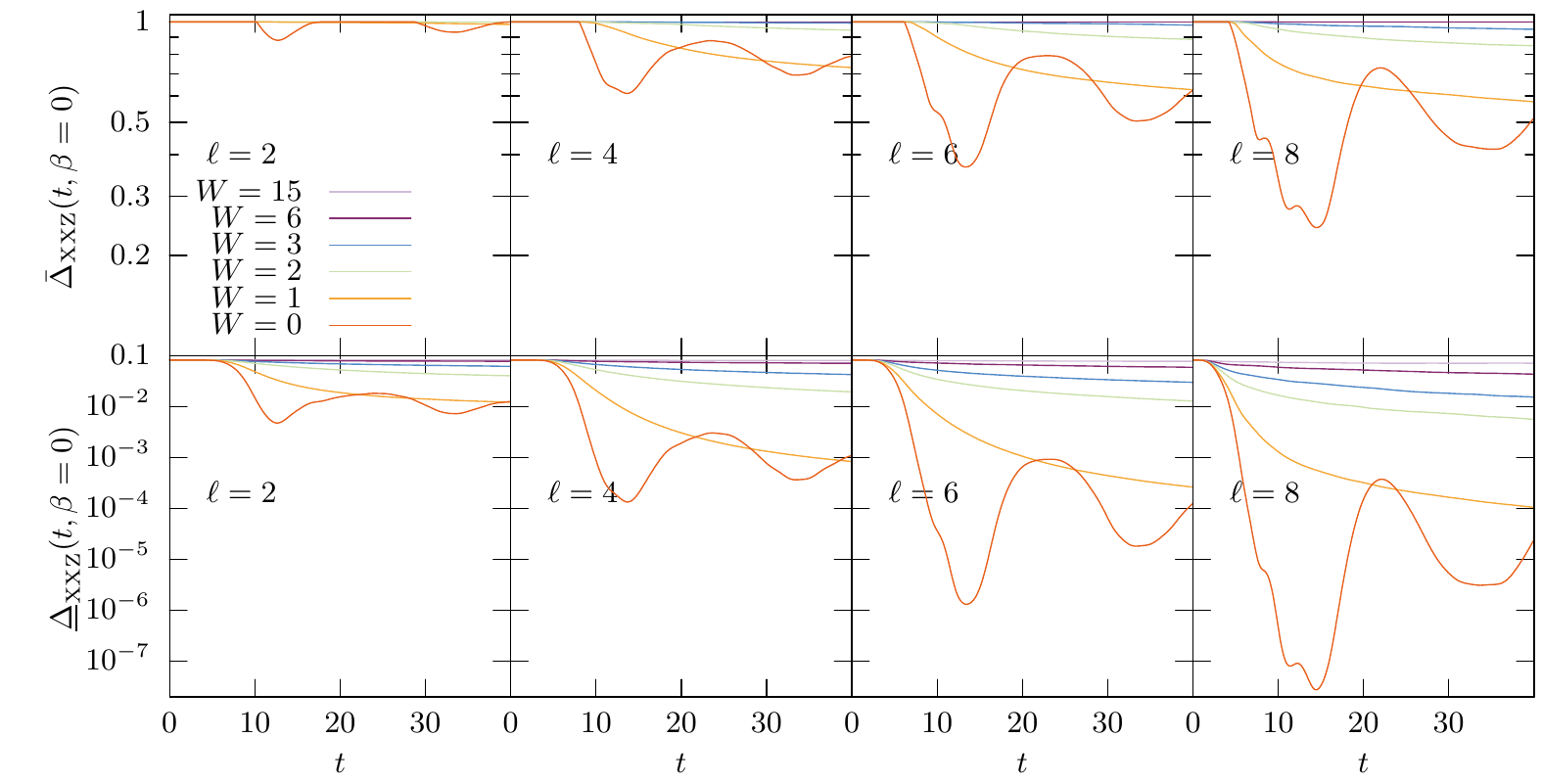}
    \caption{
    The values of $\bar{\Delta}_{\rm XXZ}$ and $\underline{\Delta}_{\rm XXZ}$ for the XXZ model with $J_z=1$, $N=12$, $\beta=0$ for various values of $W$ and $\ell=2, 4, 6, 8$.    
    Averages over 16 samples are plotted for $W>0$.
    }
    \label{fig:XXZ_l1-4-7-10}
\end{figure}

In Fig.~\ref{fig:XY_model}, we plot the time dependence of $\bar{\Delta}_{\rm XY}$ and $\underline{\Delta}_{\rm XY}$ for the XY model ($J=0$) without random magnetic field.
We set $k=1$.
Although there exists a gap between $\bar{\Delta}_{\rm XY}$ and $\underline{\Delta}_{\rm XY}$, both of them varies in a similar manner as time $t$ increases.
The plots at $\beta=0$ for different $N=8,10,12$ are qualitatively similar to each other, and they do not converge to a constant because we are here working with a single, integrable Hamiltonian for each $N$.
Introducing finite temperature, $\beta = 10$, increases the value of $\bar{\Delta}$, reflecting the decrease of the effective dimension of the Hilbert space.

The time dependence of $\bar{\Delta}_{\rm XXZ}$ and $\underline{\Delta}_{\rm XXZ}$ for the XXZ model with $J=1$ are given in Fig.~\ref{fig:XXZ_W0-1}. The two quantities decay similarly.
For $W=0$, the Hamiltonian reduces to the Heisenberg model. As it is integrable, the values are not converging. For $W=1$, the average over $16$ samples is plotted.
The late-time behavior of these averaged values of $\bar{\Delta}_{\rm XXZ}$ and $\underline{\Delta}_{\rm XXZ}$ are smoother compared to the $W=0$ case.
The value decreases as $\ell$ increases. For $\ell=9=N-1$, we always obtain $\bar{\Delta}_{\rm XXZ}=0$, which is because the system is in the $S^z=0$ subspace.

In Fig.~\ref{fig:XXZ_l1-4-7-10}, the sample averages of $\bar{\Delta}_{\rm XXZ}$ and $\underline{\Delta}_{\rm XXZ}$ for various values of $\ell$ and $W$ are provided. The upper and lower bounds again show similar time dependence.
We observe that $\bar{\Delta}_{\rm XXZ}$ monotonically increases and approaches unity as $W$ increases. This is naturally expected since the system shows MBL for large $W$, which should prevent the quantum information initially localized in the subsystem $A$ from spreading over the whole system $S$.
It is worth noting that, even though a small $W$ introduces a chaotic behavior to the system~\cite{Pal_Huse_PhysRevB.82.174411.2010,Luitz-Laflorencie-Alet-PhysRevB.91.081103}, $\bar{\Delta}_{\rm XXZ}$ remains at high values, indicating the failure of information recovery.

\begin{figure}
    \centering
    \includegraphics[width=16cm]{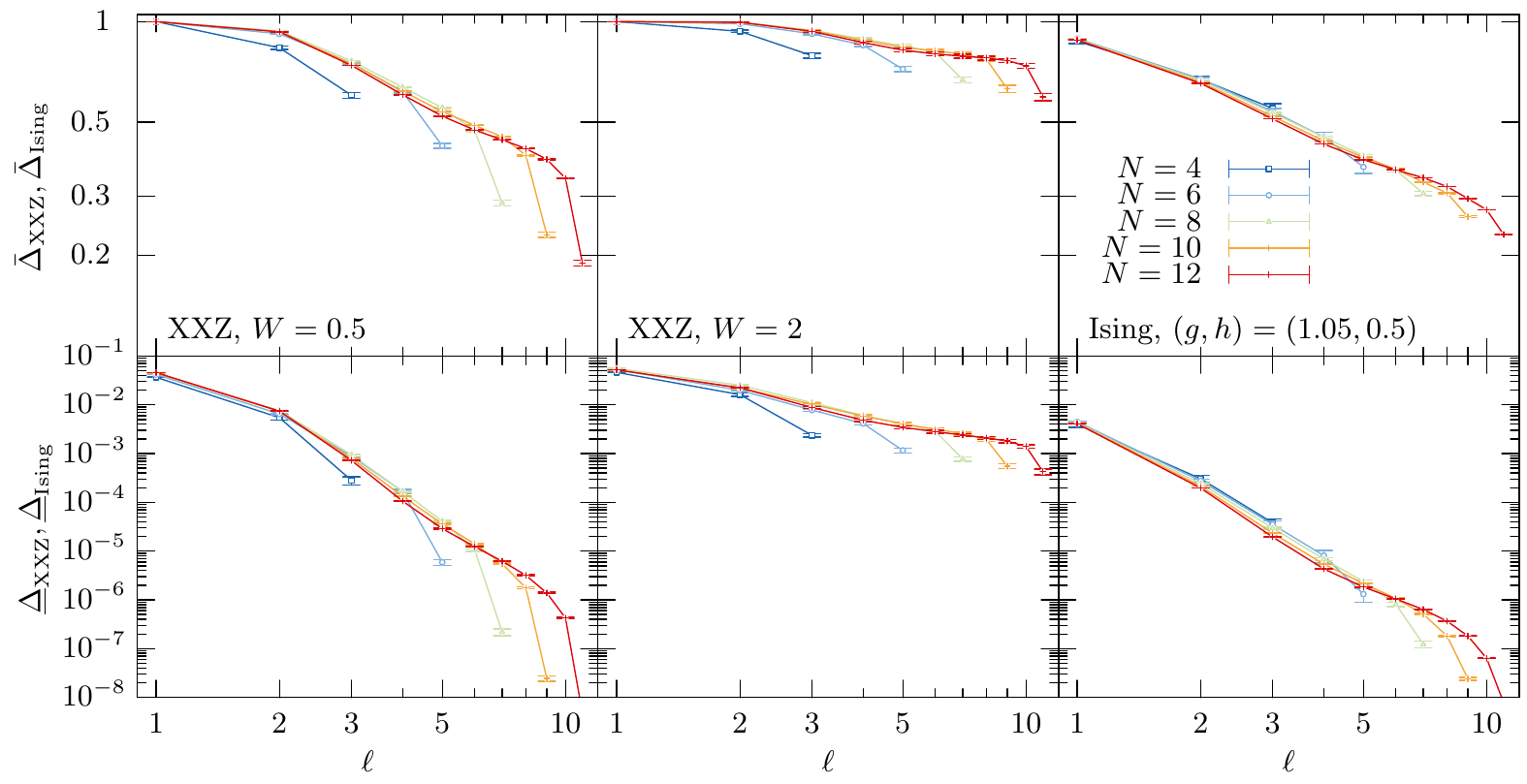}
    \caption{
    The late-time values of $\bar{\Delta}_\mathrm{XXZ}$ and $\bar{\Delta}_\mathrm{Ising}$ (top), and $\underline{\Delta}_\mathrm{XXZ}$ and $\underline{\Delta}_\mathrm{Ising}$ (bottom), plotted against $\ell$ for $N=4,6,8,10,12$ and $\beta=0$. Plots in the left (middle) column are for the XXZ model with $W=0.5$ ($W=2$), and plots in the right column are for the Ising model with $(g,h)=(1.05,0.5)$.}
    \label{fig:late-time-ell}
\end{figure}

In Fig.~\ref{fig:late-time-ell}, we plot the late-time values of $\bar{\Delta}_\mathrm{XXZ}$ and $\underline{\Delta}_{\rm XXZ}$ against $\ell$ for $\beta=0$, $W=0.5, 2$ and for various $N$. The plots almost overlap for all $N$ except when $\ell = N-1$.
Although our numerical analysis is for $N$ up to $12$, which forces $\ell$ to be at most $12$ as $\ell \leq N$, this result indicates that we can infer the values of $\bar{\Delta}_\mathrm{XXZ}$ and $\underline{\Delta}_{\rm XXZ}$ for larger $\ell$ simply by extrapolation. As both in Fig.~\ref{fig:late-time-ell} seem to decay inverse-polynomially or more slowly, we may reasonably conclude that the recovery error decays similarly as $\ell$ increases, that is, $\Delta_\mathrm{XXZ} = \Omega(1/{\rm poly}(\ell))$ unless $\ell \approx N$.

\section{Ising model with uniform magnetic field}\label{S:Ising}
A translationally invariant spin chain with nearest-neighbor Ising-type interaction and magnetic field,
\begin{align}
    \hat{H}_\mathrm{Ising}
    = -\sum_{j=1}^{N-1} \left(Z_j Z_{j+1}\right) - g\sum_{j=1}^N X_j - h\sum_{j=1}^N Z_j
    = -4\sum_{j=1}^{N-1} \left(S_{j}^zS_{j+1}^z\right)    -2g\sum_{j=1}^N S_j^x    -2h\sum_{j=1}^N S_j^z,\label{eqn:HIsing}
\end{align}
is exactly solvable if $g=0$ or $h=0$. For other choices of $(g,h)$, the model is non-integrable, and the case with $(g,h)=(1.05,0.5)$ \cite{Banuls2011} has been often studied as a prototypical model of chaotic spin chain.

\begin{figure}
    \centering
    \includegraphics[width=7.5cm]{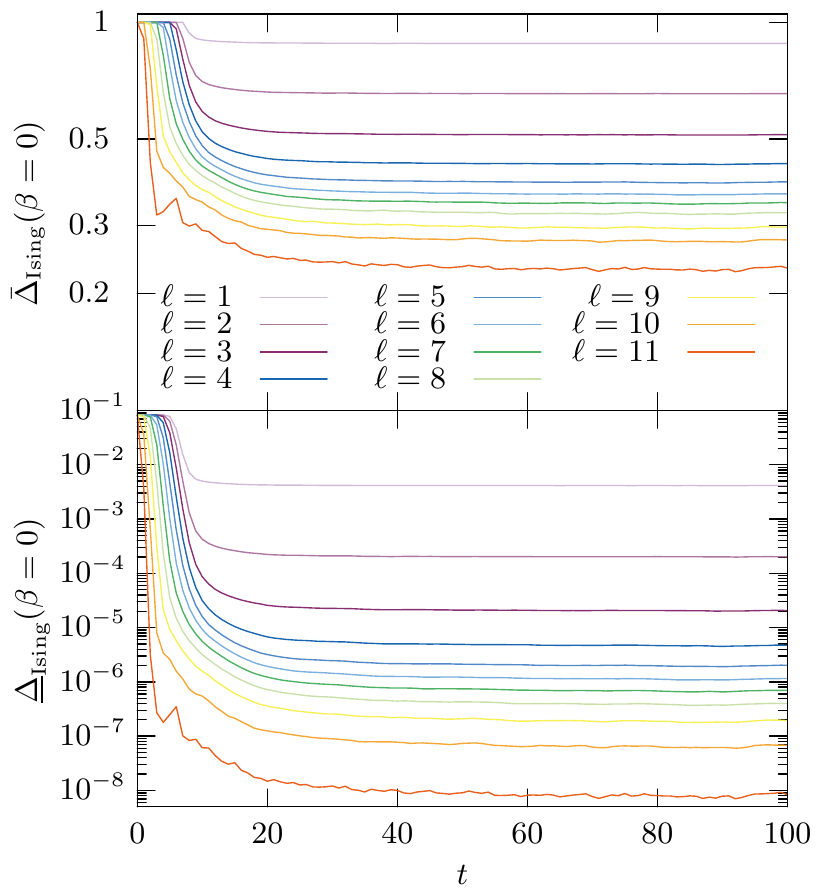}
    \caption{
    The values of $\bar{\Delta}_\mathrm{Ising}$ and $\underline{\Delta}_\mathrm{Ising}$ for the Ising spin chain with open boundary condition, $N=12$, $(g,h)=(1.05,0.5)$.}
    \label{fig:Ising}
\end{figure}

In Fig.~\ref{fig:Ising}, we plot the values of $\bar{\Delta}_\mathrm{Ising}$ and $\underline{\Delta}_\mathrm{Ising}$ for $\beta=0$ and various $\ell$. We observe that both decrease slowly as time increases and are likely to converge at the late time.
We have also checked finite $\beta$. The values of $\bar{\Delta}_\mathrm{Ising}$ and $\underline{\Delta}_\mathrm{Ising}$ are generally larger, indicating less efficient error correction.

For the late-time values of $\bar{\Delta}_\mathrm{Ising}$ and $\underline{\Delta}_\mathrm{Ising}$ against $\ell$ for several values of $N$, see Fig.~\ref{fig:late-time-ell}. Similarly to the XXZ case, the plots almost overlap except for $\ell = N-1$, and the values decay inverse-polynomially or more slowly as $\ell$ increases, except $\ell = N-1$. Thus, we may again reasonably conclude that $\Delta_\mathrm{Ising}= \Omega(1/{\rm poly}(\ell))$ unless $\ell \approx N$.

\section{Pure SYK$_4$} \label{S:pureSYK}
Here, we study the Hayden-Preskill recovery in $\mathrm{SYK}_4$ in detail. The original SYK model with all-to-all four fermion interactions among $2\ND$ Majorana fermions is henceforth denoted as the $\mathrm{SYK}_4$ model. The Hamiltonian is
\begin{equation}
    \hat{H}_{\mathrm{SYK}_4} = \sum_{1\leq a_1<a_2<a_3<a_4\leq2\ND} J_{a_1a_2a_3a_4} \hat{\psi}_{a_1}\hat{\psi}_{a_2}\hat{\psi}_{a_3}\hat{\psi}_{a_4}, \tag{\ref{eqn:SYK4}}
\end{equation}
in which the $2\ND$ Majorana fermions $\{\hat{\psi}_{j}\}_{j=1}^{2\ND}$ satisfy the anti-commutation relation
$\{\hat{\psi}_j,\hat{\psi}_k\}=\hat{\psi}_j\hat{\psi}_k+\hat{\psi}_k\hat{\psi}_j=2\delta_{jk}$, in which $\delta_{ij}$ is the Kronecker delta.
The time-independent real couplings $J_{a_1a_2a_3a_4}$ obey the Gaussian distribution $P(\{J_{a_1a_2a_3a_4}\})=\exp[-J_{a_1a_2a_3a_4}^2/2\sigma^2]/\sqrt{2\pi\sigma^2}$ with the variance $\sigma^2$.
As $\Tr[\hat{H}_{\mathrm{SYK}_4}^2]=\sum_{1\leq a_1<a_2<a_3<a_4\leq2N} J_{a_1a_2a_3a_4}^2$, we choose
\begin{align}
\sigma^2= \binom{2\ND}{4}^{-1}=\frac{12}{N(2\ND-1)(2\ND-2)(2\ND-3)},
\end{align}
so as to set the variance of the many-body energy eigenvalue to unity.
The corresponding recovery error is denoted by $\bar{\Delta}_{\mathrm{SYK}_4}(t, \beta)$.

\subsection{Effectively canceling the parity symmetry}
The SYK$_4$ model has symmetry, which divides the system into odd and even parity sectors. Accordingly, the time evolution operator is decomposed as
\begin{equation}
    \hat{U}_{{\rm SYK}_4}(t) = \hat{U}_{{\rm SYK}_4, {\rm odd}}(t) \oplus \hat{U}_{{\rm SYK}_4, {\rm even}}(t), \label{Eq:decompSYK}
\end{equation}
where $\hat{U}_{{\rm SYK}_4, p}(t)$ ($p={\rm even}, {\rm odd}$) is the unitary acting only on the sector with parity $p$.
Since the presence of symmetry is known to induce drastic changes in the Hayden-Preskill recovery~\cite{L2020,Y2019,NWK2022,https://doi.org/10.48550/arxiv.2103.01876}, which we would like to ignore in this analysis, we provide a slight modification that allows us to effectively cancel the effect of symmetry.

The modification is to embed the system $A$ of $k$ qubits into an even-parity sector of a larger system $A'$, whose Hilbert space is $\cH^{A'} := \cH^a \otimes \cH^A$ with $\cH^a = {\rm span}\{ \ket{o}, \ket{e} \}$ being a two-dimensional space that labels the parity in $A'$.
For instance, when $k=1$, we may choose
\begin{equation}
    \text{$\ket{e}^a \otimes \ket{0}^A = \ket{00}^{A'}$, $\ket{e}^a \otimes \ket{1}^A = \ket{11}^{A'}$, $\ket{o}^a \otimes \ket{0}^A = \ket{01}^{A'}$, and $\ket{o}^a \otimes \ket{1}^A = \ket{10}^{A'}$.}
\end{equation}
The maximally entangled state between $A$ and $R$ is embedded to $\ket{\Phi}^{A'R} := \ket{e}^a \otimes \ket{\Phi}^{AR}$, which is a pure state in the $2^{2k+1}$-dimensional system $A'$ with Schmidt rank $2^k$. 

Embedding $A$ into $A'$ enlarges the whole system $S$ to $S':=A' B$ of $\ND=N+1$ qubits, which we similarly decompose as $S'=sS$, where $s$ is the two-dimensional space for labeling the parity of $S$.
The time-evolution operator in $S'$ by $\hat{U}_{{\rm SYK}_4}^{S'}(t)$ is then given by 
\begin{equation}
    \hat{U}_{{\rm SYK}_4}^{S'}(t) = \ketbra{o}{o}^s \otimes \hat{U}_{{\rm SYK}_4, {\rm odd}}^{S}(t) + \ketbra{e}{e}^s \otimes \hat{U}_{{\rm SYK}_4, {\rm even}}^{S}(t).
\end{equation}
Assuming that the initial thermal state $\xi(\beta)^{B}$ is in the even-parity sector, which is $2^{N-k-1}$ dimensional, the state after the time evolution is 
\begin{align}
    \Psi^{S'R}(t, \beta) &:= \hat{U}^{S'}_{{\rm SYK}_4}(t) ( \Phi^{A'R} \otimes \xi(\beta)^{B})\hat{U}^{S' \dagger}_{{\rm SYK}_4}(t) \nonumber\\
    &= \ketbra{e}{e}^s \otimes \hat{U}^{S}_{{\rm SYK}_4, {\rm even}}(t) \bigl( \Phi^{AR} \otimes \xi(\beta)^{B}\bigr)\hat{U}^{S \dagger}_{{\rm SYK}_4, {\rm even}}(t).
\end{align}
We further assume that, in the decoding process, the information that the support of the state $\Psi^{S'R}(t, \beta)$ is in the even-parity sector of $\cH^{S'}$ is available. Then the state relevant to the degree of decoupling is the one in which $s$ and $C$ are traced over:
\begin{align}
    \Psi^{DR}(t, \beta) 
    &= \Tr_{s,C} \bigl[ \Psi^{S'R}(t, \beta) \bigr] \nonumber\\
    &= \Tr_{C} \bigl[ \hat{U}^{S}_{{\rm SYK}_4, {\rm even}}(t) ( \Phi^{AR} \otimes \xi(\beta)^{B})\hat{U}^{S \dagger}_{{\rm SYK}_4, {\rm even}}(t) \bigr]. \label{eer}
\end{align}
In this way, the effect of symmetry of the SYK$_4$ model is effectively canceled.

It should be emphasized that this modification leads to a slight change in the degree of decoupling, resulting in a slight change of the upper bound $\bar{\Delta}$ on the recovery error. To see this, we consider the degree $\Sigma'_{\rm Haar}$ of decoupling when $\hat{U}^{S}_{{\rm SYK}_4, {\rm even}}(t)$ in Eq.~\eqref{eer} is replaced with the Haar random unitary $U^S_{\rm Haar}$ acting merely on the even-parity sector with dimension $2^{N-k-1}$. 
A straightforward calculation leads to
\begin{equation}
\log[ 2 \Sigma'_{\rm Haar}(\beta)] \leq \frac{N+k-H(\beta)-1}{2} - \ell.
\end{equation}
This leads to
\begin{equation}
\Delta'_{\rm Haar}(\beta)
\leq
2^{\frac{1}{2} \bigl( \frac{N+k-H(\beta)-1}{2} - \ell \bigr)} = 2^{\frac{1}{2} \bigl( \frac{N+k-H(\beta)}{2} - \ell \bigr)-\frac{1}{4}}.\label{eqn:paritySYK4error}
\end{equation}
This differs from Eq.~\eqref{Eq:Haar} by factor $2^{-\frac{1}{4}}$, which arises from the fact that the information about the parity sector is available in the decoding process. Note also that, since we have restricted the initial thermal state $\xi(\beta)^B$ on the even-parity sector, its entropy also changes. For instance, its maximum value is $N-k-1$ rather than $N-k$.

\subsection{Recovery errors in the $\mathrm{SYK}_4$ model}
\begin{figure*}
    \centering
    \includegraphics[]{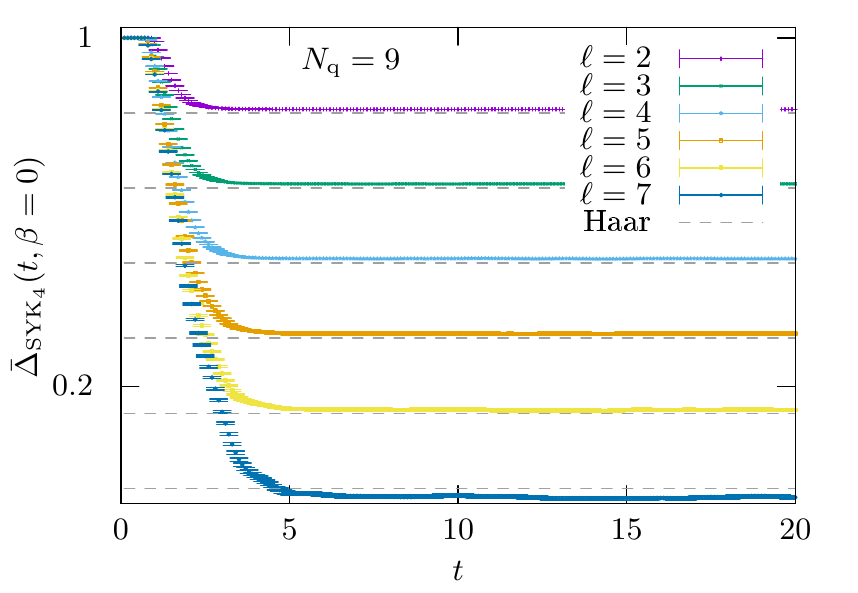}
    \includegraphics[]{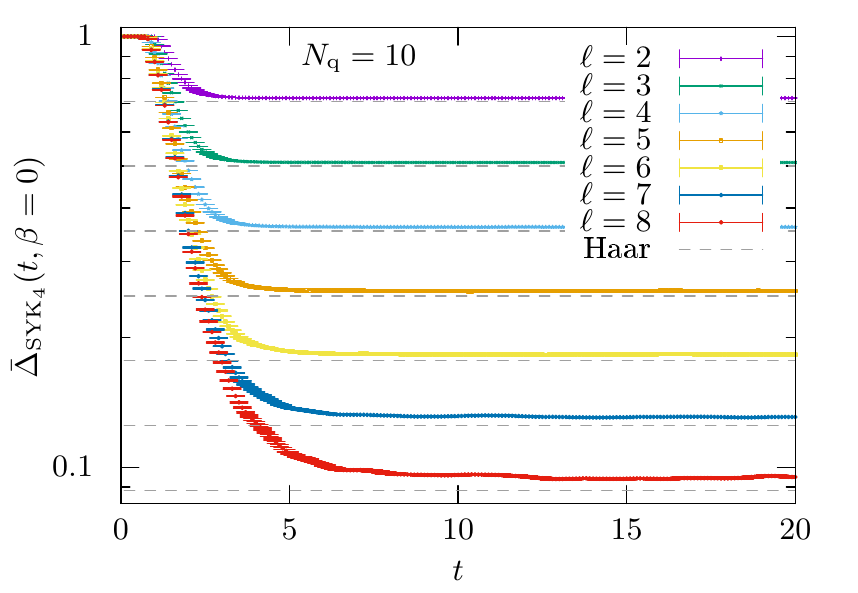}
    \includegraphics[]{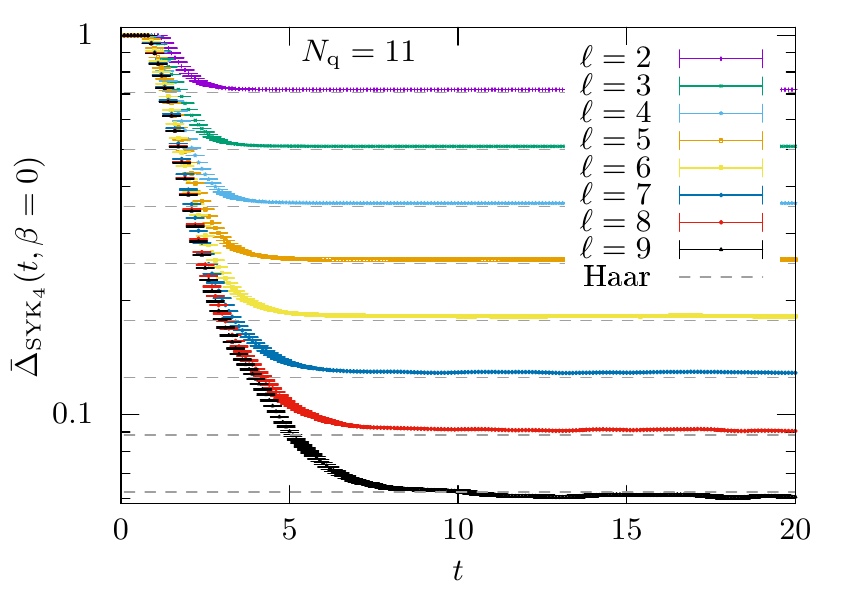}
    \includegraphics[]{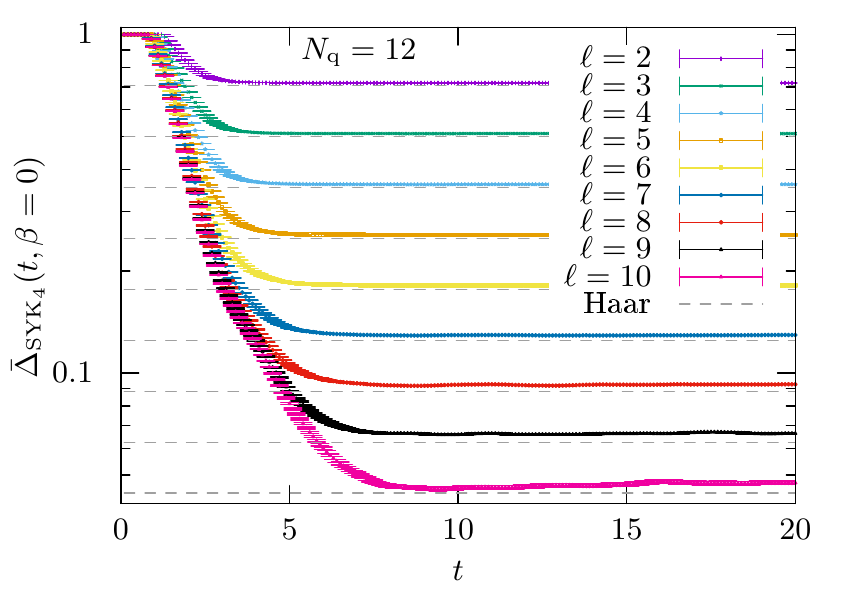}
    \caption{
    Semilogarithmic plot of the value of $\bar{\Delta}_{\hat{H}_{\mathrm{SYK}_4}}(t,\beta=0)$ against $t$ for $\ND=9,10,11,12$ and $2\leq l < \ND-k$. Average over $2^{19-\ND}$ samples is taken.
    For $\ell=1$, $\bar{\Delta}_{\hat{H}_{\mathrm{SYK}_4}}(t,\beta=0)$ is $\sim 1$ for all $t$.
    The dashed lines represent $\bar{\Delta}_{\rm Haar}(\beta = 0)=2^{\frac{1}{2}(1 - \ell)}$ given in Eq.~\eqref{Eq:Haar} for $\ell=2,3,\ldots,\ND-2$.
    See Fig.~\ref{fig:k1N13} for the plot for $\ND=13$.}
    \label{fig:k1N10}
\end{figure*}

The collision entropy of $\xi_B(\beta)$ is
\begin{equation}
    H(\beta) = -\log \Tr[\rho(\beta)^2]
    = -\log \Tr \frac{[e^{-2\beta H}]}{Z(\beta)^2}
    = -\log \frac{Z(2\beta)}{Z(\beta)^2}
    = 2\log Z(\beta) - \log Z(2\beta)
    = \frac{1}{\ln 2}(F(2\beta) - 2F(\beta)),
\end{equation}
in which $F(\beta)=-\ln Z(\beta)$ is the Helmholtz free energy.
For $\beta=0$, $Z(\beta=0)$ is simply the dimension of the Hilbert space of $B$, and we have $H(\beta=0)=\log Z(\beta=0)=N-k-1$. Therefore,
the right-hand side of \eqref{eqn:paritySYK4error} is
\begin{equation}
2^{\frac{1}{2} \bigl( \frac{N+k-(N-k-1)}{2} - \ell \bigr)-\frac{1}{4}}
=2^{\frac{k-\ell}{2}};\ 
\bar{\Delta}'_{\rm Haar}(\beta=0)=\min \{1, 2^{\frac{k-\ell}{2}}\}.
\end{equation}

\begin{figure*}
    \centering
    \includegraphics[width=0.32\linewidth]{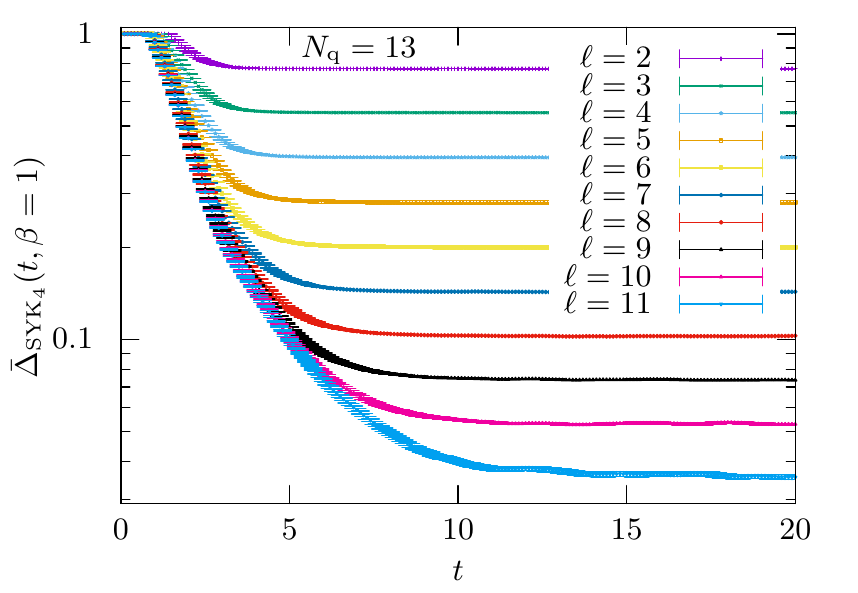}
    \includegraphics[width=0.32\linewidth]{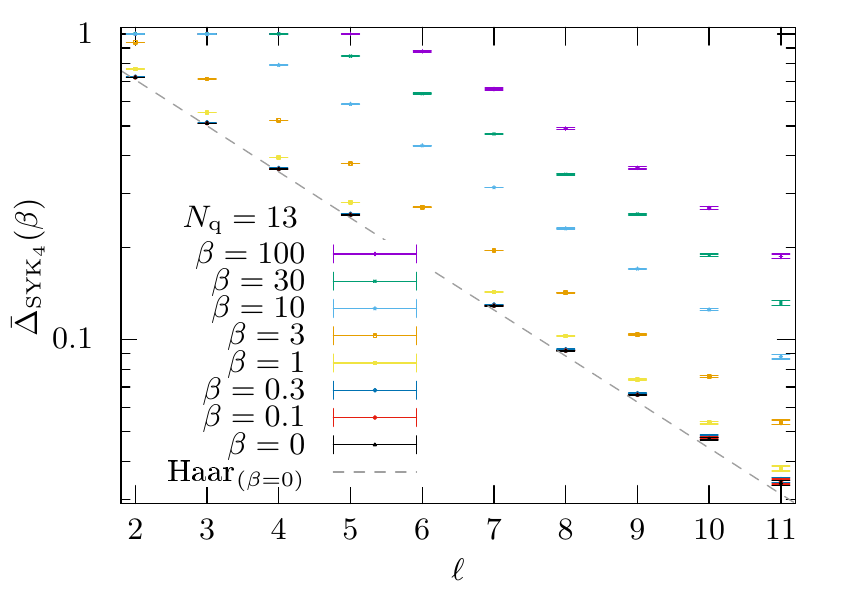}
    \includegraphics[width=0.32\linewidth]{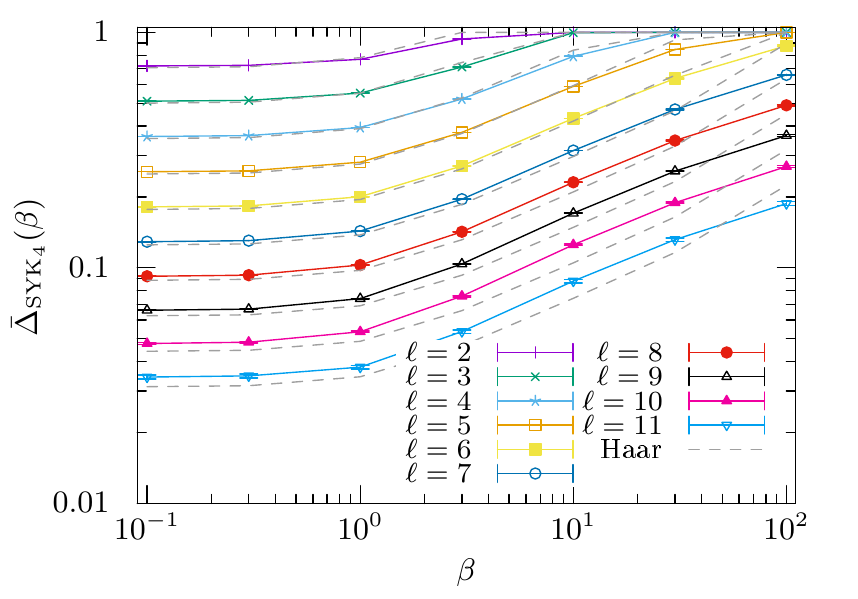}
    \caption{
    Left: Semilogrithmic plot of the value of $\bar{\Delta}_{\hat{H}_{\mathrm{SYK_4}}}(t,\beta=1)$ against $t$ for $\ND=13$.
    For $\beta=0$, see Fig.~\ref{fig:k1N13} in the main text.
    Center: Late-time value of $\bar{\Delta}_{\hat{H}_{\mathrm{SYK_4}}}(t,\beta)$ averaged over $64$ samples plotted against $\ell$ for $\ND=13$.
    Right: Late-time value $\bar{\Delta}_{\hat{H}_{\mathrm{SYK_4}}}(t,\beta)$ averaged over $64$ samples plotted against $\beta$ for $\ND=13$. Dashed lines represent $\bar{\Delta}'_{\mathrm{Haar}}(\beta)$.
    }    \label{fig:k1N13b1}
\end{figure*}

In Fig.~\ref{fig:k1N10}, we plot upper bound $\bar{\Delta}_{\mathrm{SYK}_4}$ on the recovery error against time $t$ for various $\ell$, for $9\leq \ND\leq 12$ and infinite temperature $\beta = 0$. Since the $\mathrm{SYK}_4$ model has randomness of choosing the coupling constant $J_{a_1a_2a_3a_4}$, we take an average over many samples.
We clearly observe that all curves rapidly decay and approach the bound $\bar{\Delta}_{\rm Haar}$ for the Haar random dynamics, indicating that the Hayden-Preskill recovery is quickly achieved by the dynamics of $\mathrm{SYK}_4$. 
This is also the case for finite temperature, as observed in Fig.~\ref{fig:k1N13b1}. The late-time value of $\bar{\Delta}_{\mathrm{SYK}_4}$ is in proportion to $2^{-H(\beta)/4}$ as expected from \eqref{eqn:paritySYK4error} for $\beta\lesssim10$, beyond which small deviations occur presumably due to the small number of eigenenergies within $1/\beta$ of the ground state.

Note that the $\mathrm{SYK}_4$ model has a $\ND \mod 4$ periodicity, depending on which the eigenenergy statistics resembles the Gaussian Unitary, Orthogonal, and Symplectic Ensembles (GUE, GOE, and GSE). We have checked that this periodicity does not affect much on the Hayden-Preskill recovery, implying that the Hayden-Preskill recovery is achieved by those ensembles of random Hermitian matrices. This might be of interest since the dynamics generated by GUE, GOE, and GSE is not Haar random, but achieves the Hayden-Preskill recovery that was shown based on a Haar random unitary.

\begin{figure}
    \centering
    \includegraphics[]{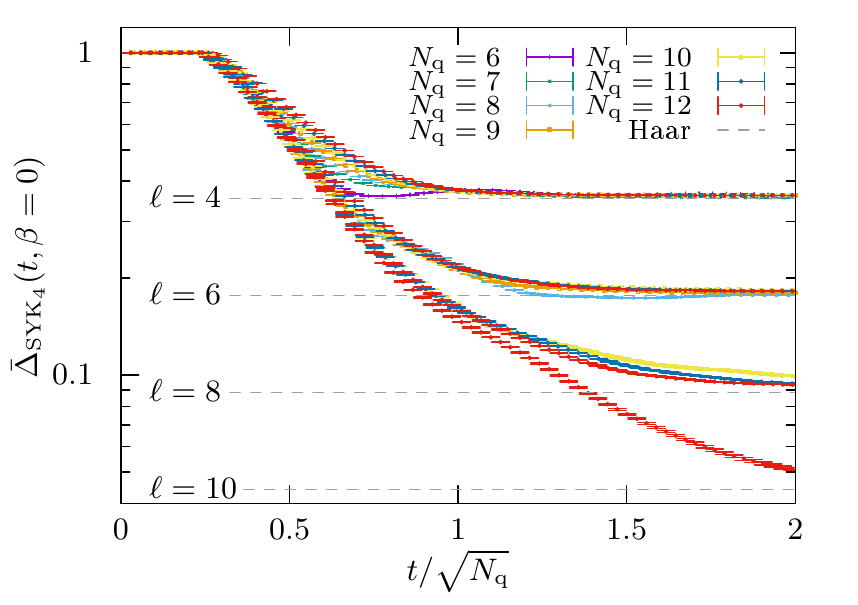}
    \caption{
    $\bar{\Delta}_{\hat{H}_{\mathrm{SYK}_4}}(t,\beta=0)$ plotted against $t/\sqrt{\ND}$, for $\ell=4,6,8,10$ from top to bottom.
    Average over $2^{19-\ND}$ samples is taken. 
    The curves show better agreement as $\ell$ is increased, which is not observed for other choices of the exponent of $\ND$ for scaling.}
    \label{fig:SYK4scaledTime}
\end{figure}

\subsection{Convergence time}
To check the time scale needed for $\bar{\Delta}_{\mathrm{SYK}_4}$ converging to $\bar{\Delta}_{\rm Haar}$, we plot in Fig.~\ref{fig:SYK4scaledTime} $\bar{\Delta}(t,\beta=0)$ for $\ell=4,6,8,10$ against re-scaled time $t/\sqrt{\ND}$. The plot indicates that the convergence time is likely to be $\sqrt{\ND}$.

\section{Sparse SYK$_4$}\label{S:sparseSYK}
We consider the sparse $\mathrm{SYK}_4$ model~\cite{xu2020sparse,Garc_a_Garc_a_2021,SparsePM1SYK}, referred to as $\mathrm{spSYK}_4$. The model has a parameter $\Kcpl$ that counts the number of non-zero coupling constant $J_{a_1a_2a_3a_4}$ in Eq.~\eqref{eqn:SYK4}. Other than that, the Hamiltonian of $\mathrm{spSYK}_4$ is defined similarly to that of $\mathrm{SYK}_4$.
When $\Kcpl = \binom{\ND}{4}$, then $\mathrm{spSYK}_4$ reduces to $\mathrm{SYK}_4$.
Note that this approach to reducing the number of non-zero parameter is drastically different from the approach in \cite{Dumitriu_2002}, where tridiagonal matrix models generalizing the Gaussian and Wishart dense matrix models were introduced. The $\mathrm{SYK}_4$ Hamiltonian is still sparse when presented as a matrix in the many-body Hilbert space.

For simpliity, we further introduce a constraint that a half of the non-zero couplings is set to $1/\sqrt{\Kcpl}$ and the other half is $-1/\sqrt{\Kcpl}$, which we call $\pm \mathrm{spSYK}_4$ model.
The corresponding upper bound of the recovery error is denoted by $\bar{\Delta}_{\pm \mathrm{spSYK}_4}(t, \beta)$ for time $t$ and the inverse temperature $\beta$.

The sparse $\mathrm{SYK}_4$ model is known to have a transition from quantum chaos to integrability by varying the number $\Kcpl$ of non-zero coupling constant.
For $\Kcpl\ll 2\ND$, we typically have extra degeneracy in the spectrum than expected from the symmetry, and the spectral statistics of the distinct eigenvalues is not random-matrix like.
For $\Kcpl\gtrsim 2\ND$, such extra degeneracy disappears for practically all samples, and the spectral statistics strongly resembles that of the Gaussian random matrix ensemble with the corresponding symmetry \cite{xu2020sparse,Garc_a_Garc_a_2021,SparsePM1SYK}.
 It is surprising to some extent that the $\pm \mathrm{spSYK}_4$ recovers important properties of $\mathrm{SYK}_4$ when $\Kcpl = O(N)$, which is by far smaller than $\Kcpl = \binom{\ND}{4}$, which is needed for $\hat{H}_{\mathrm{spSYK}_4}$ reducing to $\hat{H}_{\mathrm{SYK}_4}$.

\begin{figure}
    \centering
    \includegraphics[width=0.32\linewidth]{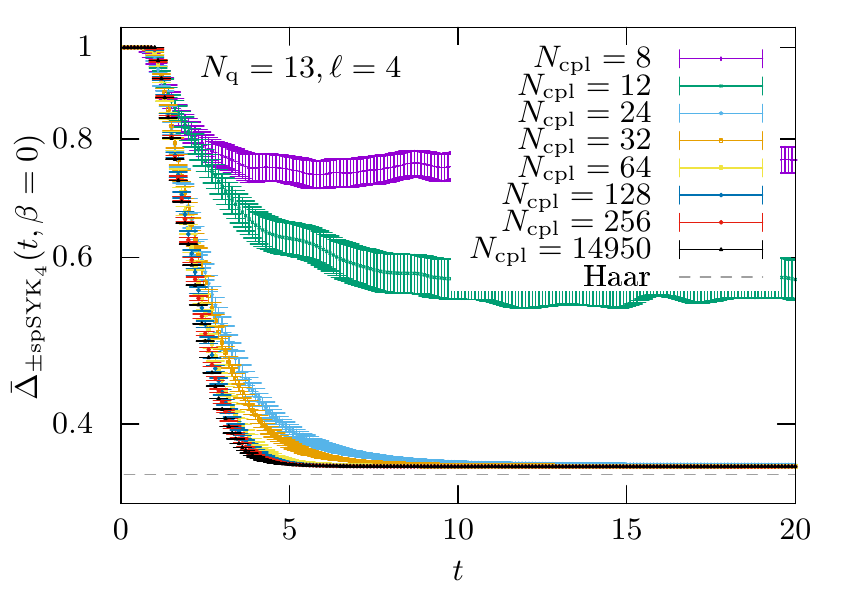}
    \includegraphics[width=0.32\linewidth]{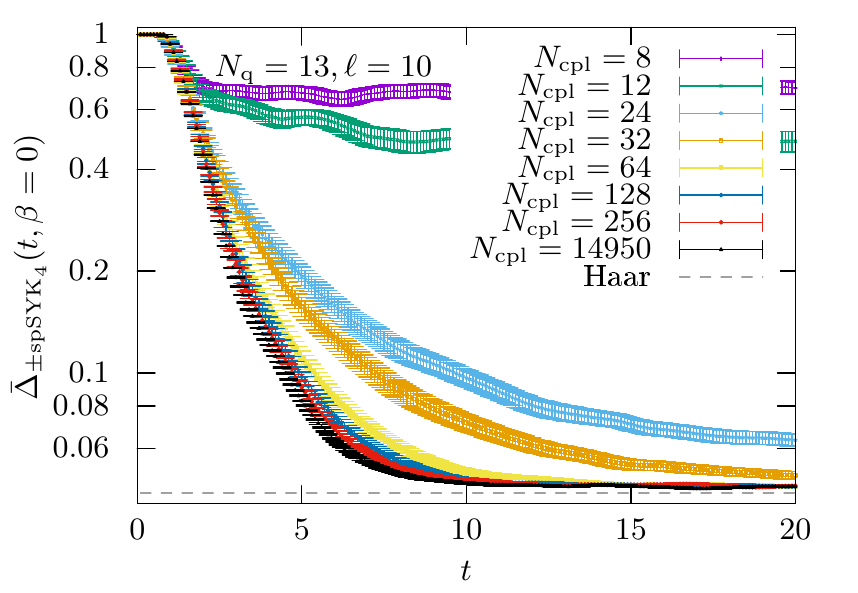}
    \includegraphics[width=0.32\linewidth]{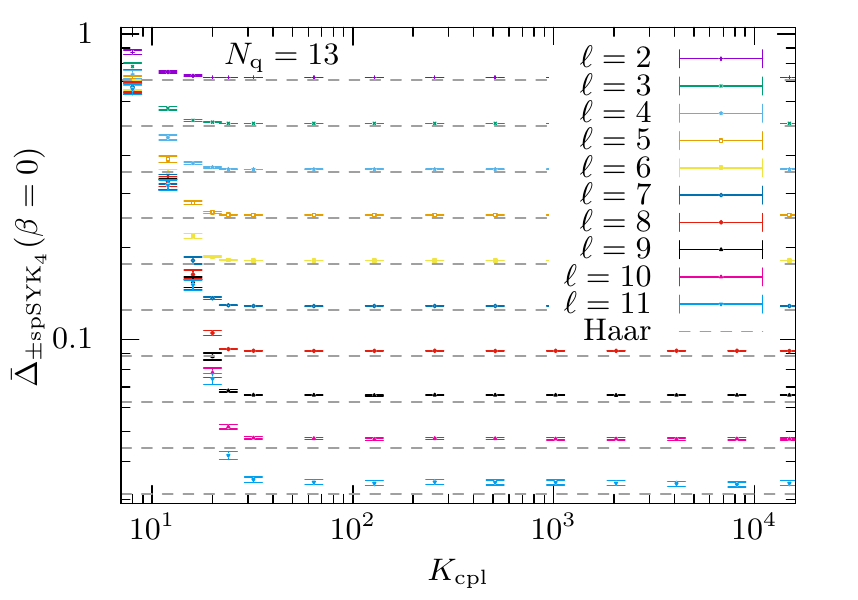}
    \caption{
    The value of $\bar{\Delta}_{\pm \mathrm{spSYK}_4}(t,\beta=0)$ plotted against $t$ for $\ND=13$ and $\ell=4$ (left), $\ell=10$ (center). 64 samples are used. The result for the $\mathrm{SYK}_4$ model as well as the Haar value, $\bar{\Delta}_{\rm Haar}(\beta = 0) = 2^\frac{1-\ell}{2}$, are also shown.
        (right) The late-time value of $\bar{\Delta}_{\pm \mathrm{spSYK}_4}(t,\beta=0)$ plotted for $\ND=13$, and various $\ell$ against the number $\Kcpl$ of nonzero couplings. 64 samples are used.
    }
    \label{fig:Sparsek1l-4-7}
\end{figure}

In Fig.~\ref{fig:Sparsek1l-4-7}(a--b), we plot $\bar{\Delta}_{\pm \mathrm{spSYK}_4}(t, \beta)$ for $\ell=4, 10$ against $t$ for various values of $\Kcpl$. At an earlier time, $\bar{\Delta}_{\pm \mathrm{spSYK}_4}(t, \beta)$ starts decreasing for any $\Kcpl (>0)$ as time $t$ increases, which is similar to the $\mathrm{SYK}_4$ model, but it soon witnesses the difference from $\mathrm{SYK}_4$ when $\Kcpl$ is small.
For $\Kcpl \lesssim 16$, $\bar{\Delta}_{\pm \mathrm{spSYK}_4}$ is unlikely to converge to the Haar value $\bar{\Delta}_{\rm Haar, th}$ even in the large time limit, while for large $\Kcpl$, $\bar{\Delta}_{\pm \mathrm{spSYK}_4}$ seems to eventually converge to $\bar{\Delta}_{\rm Haar, th}$ as $t$ increases.
Hence, we observe from these plots that at least two differences come up in the $\pm \mathrm{spSYK}_4$ model in comparison with the $\mathrm{SYK}_4$ model. One is the converging value of $\bar{\Delta}_{\pm \mathrm{spSYK}_4}(t = \infty, \beta)$, and the other is the time scale for convergence.

To closely investigate the converging value, we plot in Fig.~\ref{fig:Sparsek1l-4-7}(c) the late-time value of $\bar{\Delta}_{\pm \mathrm{spSYK}_4}$ for various $\ell$ against $\Kcpl$ for $2\ND=26$ Majorana fermions.
It is clear that the value is nearly identical to the Haar value for $\Kcpl \gtrsim 30$. This convergence to the dense SYK limit is consistent with the eigenenergy statistics.\cite{SparsePM1SYK}
For the density of states plotted in the left panel of Fig.~\ref{fig:sparse_SYK4+2_DoS}, we observe that while the overall shape stays similar, the fluctuation disappears as we increase $\Kcpl$.
The fluctuation observed for smaller $\Kcpl$ originates from the decrease in the number of distinct eigenvalues due to the additional degeneracy when emergent conserved quantities appear.\cite{Garc_a_Garc_a_2021,SparsePM1SYK}

\section{SYK$_{4+2}$}\label{S:SYK4+2}
Because the trace of the product of six Majorana fermions is zero when four of them differ with each other,
for the SYK$_{4+2}$ model defined in eq.~\eqref{eqn:SYK4+2}, we have
\begin{align}
    \Tr \hat{H}_{\mathrm{SYK}_{4+2}}(\theta)^2
    =(\cos^2\theta)\Tr \hat{H}_{\mathrm{SYK}_{4}}^2
    +(\sin^2\theta)\Tr \hat{H}_{\mathrm{SYK}_{2}}^2
    =2^\ND.
\end{align}
Note that the normalization in \cite{Monteiro2021PRR} is so that $\hat{H}_{\mathrm{SYK}_{4+2}}(\theta) = \hat{H}_{\mathrm{SYK}_4}+\delta\hat{H}_{\mathrm{SYK}_2}$
with the variance of the many-body eigenvalues for $\hat{H}_{\mathrm{SYK}_4}$ is $24(2\ND)^{-4}\begin{pmatrix}2\ND\\4\end{pmatrix}\sim 1$,
and the variance of the \textit{single-particle} eigenvalues for $\hat{H}_{\mathrm{SYK}_2}$ is $\delta^2(2\ND)^{-3}(2\ND-2)(2\ND-3)\sim \delta^2/(2\ND)$.

\subsection{The density of states for SYK$_{4+2}$}
\begin{figure}
    \centering
    \includegraphics[width=0.45\linewidth]{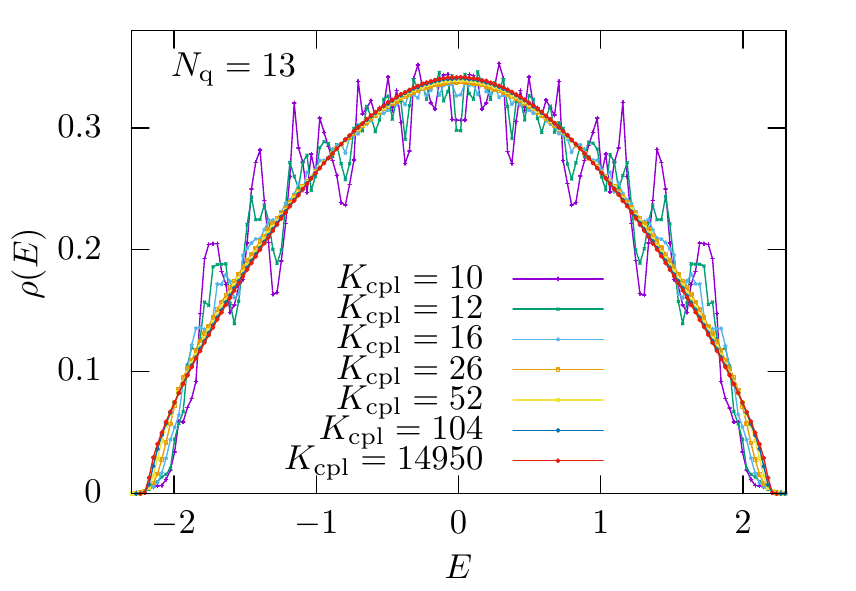}
    \includegraphics[width=0.45\linewidth]{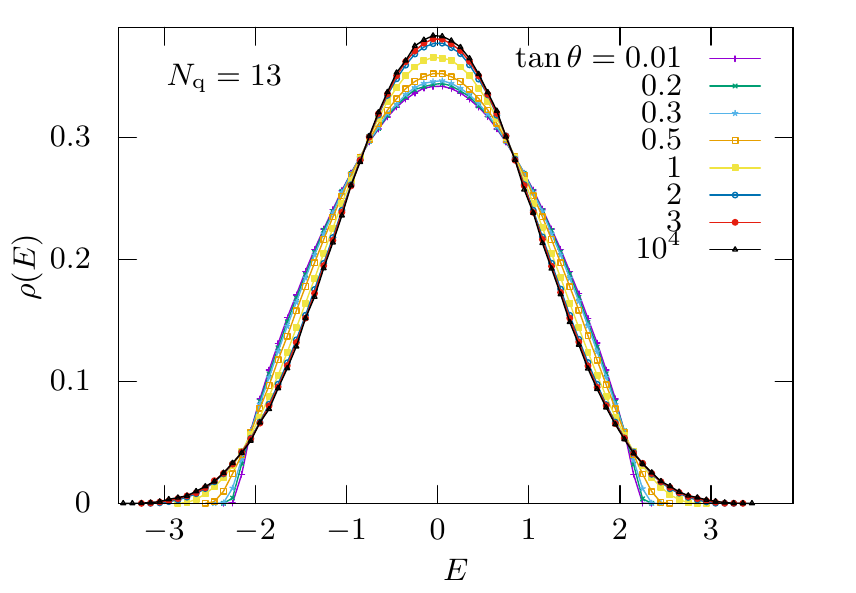}
    \caption{
    The normalized density of states for (left) $\hat{H}_{\pm \mathrm{spSYK}_4}$ plotted for $\ND=13$ and 2048 samples, and for (right) $\hat{H}_{\mathrm{SYK}_{4+2}}$ plotted for $\ND=13$ and 64 samples.
    }
    \label{fig:sparse_SYK4+2_DoS}
\end{figure}

We plot in the right panel of Fig.~\ref{fig:sparse_SYK4+2_DoS} the density of states for various values of $\tan\theta$.
The variance of the eigenstate energy is fixed at unity.
As $\tan\theta$ is increased, the peak becomes higher and the tails become thicker, changing from a RMT-like spectrum, which is a sign for quantum chaos and is the case for SYK$_4$, to a Gaussian-shape spectrum, which differs from SYK$_4$.

\subsection{Numerical results for SYK$_{4+2}$}
We here provide numerical results of the Hayden-Preskill recovery in the $\mathrm{SYK}_{4+2}$ model. The Hamiltonian is given by
\begin{align}
    \hat{H}_{\mathrm{SYK}_{4+2}}(\theta) &= (\cos\theta) \hat{H}_{\mathrm{SYK}_4} + (\sin\theta) \hat{H}_{\mathrm{SYK}_2}\nonumber\\
    &= (\cos\theta)\left(\hat{H}_{\mathrm{SYK}_4} + (\tan\theta) \hat{H}_{\mathrm{SYK}_2}\right), \tag{\ref{eqn:SYK4+2}}
\end{align}
which maps to the normalization in \cite{Monteiro2021PRR} by $\delta=\tan\theta$.
The $\mathrm{SYK}_{4+2}$ model trivially reduces to $\mathrm{SYK}_{4}$ when $\delta = 0$.
Below, we denote by $\bar{\Delta}_{\mathrm{SYK}_{4+2}}(t, \beta)$ the corresponding upper bound on the recovery error and investigate it.

\begin{figure*}
    \centering
    \includegraphics[width=0.45\linewidth]{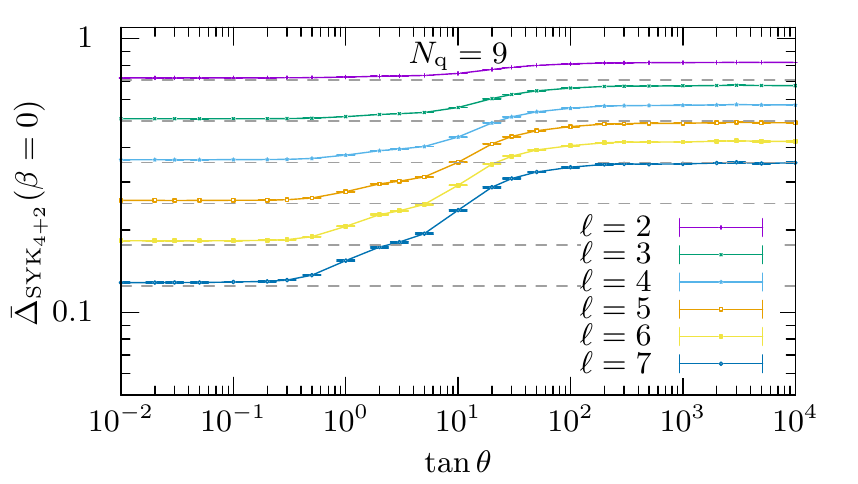}
    \includegraphics[width=0.45\linewidth]{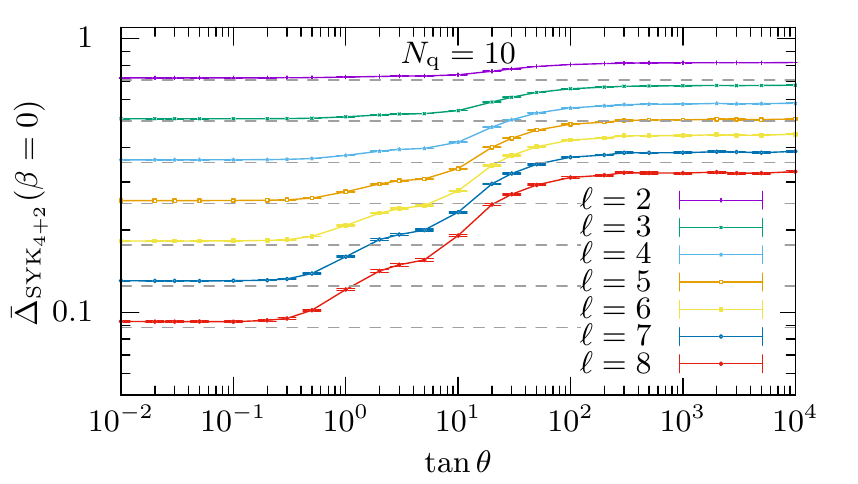}
    \includegraphics[width=0.45\linewidth]{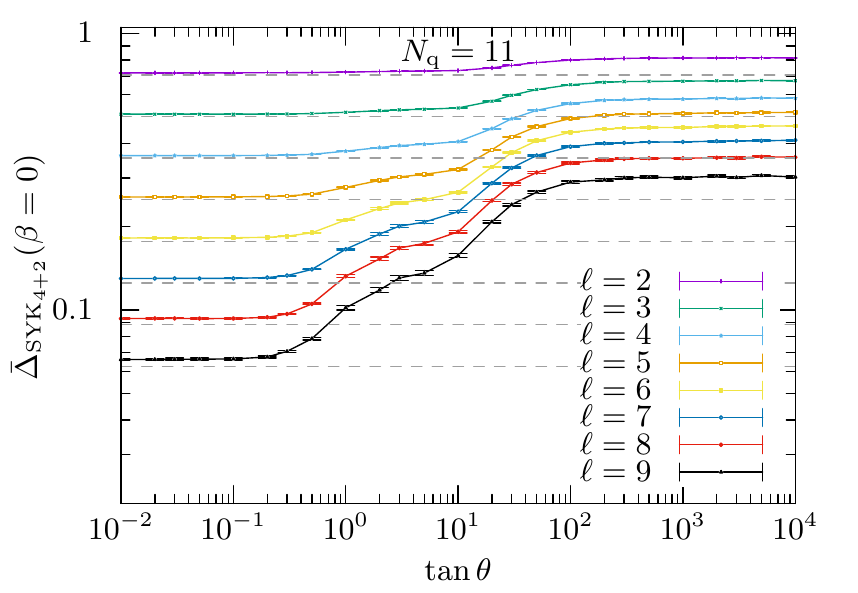}
    \includegraphics[width=0.45\linewidth]{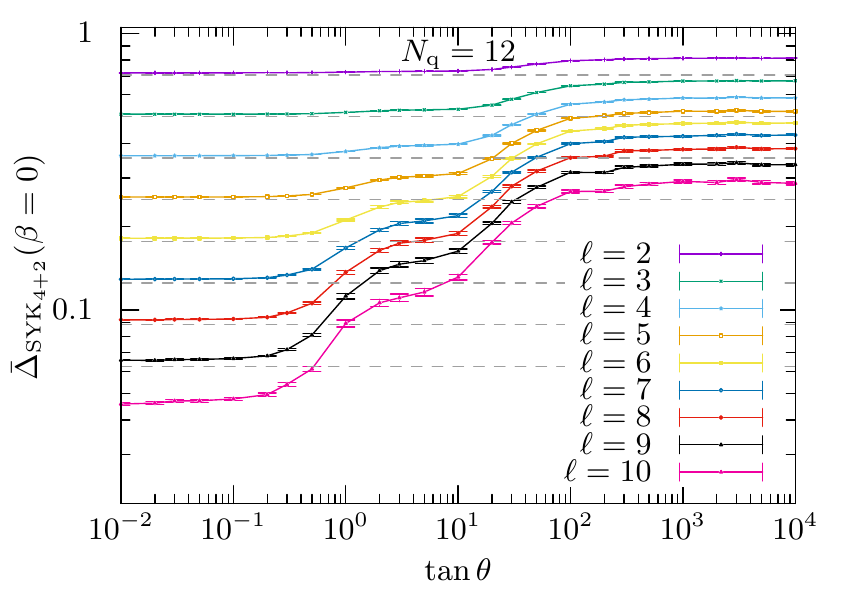}
    \caption{
    The value of $\bar{\Delta}_{\mathrm{SYK}_{4+2}}(t,\beta=0)$ plotted for various $\ell$ against the value of $\delta$ for $\ND=9,10,11,12$. $2^{19-\ND}=2^{10},2^{9},2^{8},2^{7}$ samples are used and average over the values for $t=(1,2,\ldots,10)\times 10^6$ is taken. The lines connecting the data points are guide to the eye.
    }
    \label{fig:delta_dependence_N11}
\end{figure*}

By varying the strength $\delta$ of the $\mathrm{SYK}_2$ term, this Hamiltonian shows a transition from quantum chaos, in the sense of eigenenergy statistics, to Fock space many-body localization (F-MBL)
at $\delta_\mathrm{c}\simeq \ND^2\ln \ND$ \cite{Monteiro2021PRR},
which corresponds to $\tan\theta=\mathcal{O}(10^2)$ for $9\leq\ND\leq 13$ that we numerically study in this work.

In Fig.~\ref{fig:delta_dependence_N11} we plot the late time behaviour of $\bar{\Delta}_{\mathrm{SYK}_{4+2}}(t, \beta)$ by setting $\beta=0$ as a function of the SYK$_2$ coupling strength $\delta=\tan\theta$.
There are two characteristic values of $\delta=\tan\theta$, $\delta_0=\tan\theta_0 \approx 0.1$ and $\delta_1=\tan\theta_1 \approx 3$, yielding three plateau-like shapes of $\bar{\Delta}_{\mathrm{SYK}_{4+2}}$.
When $0 \leq \theta \leq \theta_0$, $\bar{\Delta}_{\mathrm{SYK}_{4+2}} \approx \bar{\Delta}_{\rm Haar}$. This implies that the Hayden-Preskill recovery in the $\mathrm{SYK}_4$ model is stable against small perturbations by the $\mathrm{SYK}_2$ terms.
When $\theta_0 \leq \theta \leq \theta_1$, $\bar{\Delta}_{\mathrm{SYK}_{4+2}}$ starts deviating from the Haar value $\bar{\Delta}_{\rm Haar}(\beta = 0)$ and increases linearly in terms of $\delta$ until it reaches the second plateau. The second plateau ends at $\theta = \theta_1$ and, for $\theta \geq \theta_1$, $\bar{\Delta}_{\mathrm{SYK}_{4+2}}$ restarts increasing until the third plateau.

\end{document}